\documentclass[final,3p,times,sort&compress]{elsarticle}
\usepackage[colorlinks,citecolor=magenta,linkcolor=blue]{hyperref}
\usepackage{graphicx}
\usepackage{natbib}
\usepackage{amsmath,amssymb}
\usepackage{txfonts,bm,listings,xcolor}

\newcounter{bla}

\lstdefinelanguage{Julia}{morekeywords={using,
    push, open, do, for, in, eachindex,
    end, Dict, String, Any, im, printf,
    LOAD_PATH, ENV},
    sensitive=true,
    morecomment=[l]{\#},
    morestring=[b]",
}

\lstdefinelanguage{TOML}{morekeywords={BASE,
    MaxEnt, StochOM, StochAC, StochSK},
    sensitive=true,
    morecomment=[l]{\#},
    morestring=[b]",
}

\journal{Computer Physics Communications}

\begin{document}

\begin{frontmatter}

\title{ACFlow: An open source toolkit for analytical continuation of quantum Monte Carlo data}

\author[a]{Li Huang\corref{author}}

\cortext[author] {Corresponding author.\\\textit{E-mail address:} huangli@caep.cn}
\address[a]{Science and Technology on Surface Physics and Chemistry Laboratory, P.O. Box 9-35, Jiangyou 621908, China}

\begin{abstract}
The purpose of analytical continuation is to establish a real frequency spectral representation of single-particle or two-particle correlation function (such as Green's function, self-energy function, and dynamical susceptibilities) from noisy data generated in finite temperature quantum Monte Carlo simulations. It requires numerical solutions of a family of Fredholm integral equations of the first kind, which is indeed a challenging task. In this paper, an open source toolkit (dubbed ACFlow) for analytical continuation of quantum Monte Carlo data is presented. We at first give a short introduction to the analytical continuation problem. Next, three primary analytical continuation algorithms, including maximum entropy method, stochastic analytical continuation, and stochastic optimization method, as implemented in this toolkit are reviewed. And then we elaborate major features, implementation details, and basic usage of this toolkit. Finally, four representative examples are shown to demonstrate usefulness and flexibility of the ACFlow toolkit. 
\end{abstract}


\end{frontmatter}

\noindent {\bf PROGRAM SUMMARY}\\

\begin{small}
\noindent
{\em Program Title:} ACFlow \\
{\em CPC Library link to program files:} (to be added by Technical Editor) \\
{\em Developer's repository link:} https://github.com/huangli712/ACFlow \\
{\em Code Ocean capsule:} (to be added by Technical Editor)\\
{\em Licensing provisions (please choose one):} GPLv3 \\
{\em Programming language:} Julia \\
{\em Supplementary material:} \\
{\em Journal reference of previous version:}* \\
{\em Does the new version supersede the previous version?:}* \\
{\em Reasons for the new version:}* \\
{\em Summary of revisions:}* \\
{\em Nature of problem (approx. 50-250 words):} 
Most of the quantum Monte Carlo methods work on imaginary axis. In order to extract physical observables and compare them with the experimental results, analytical continuation must be done in the post-processing stage to convert the quantum Monte Carlo simulated data from imaginary axis to real axis. \\
{\em Solution method (approx. 50-250 words):}
Three established analytical continuation methods, including maximum entropy method, stochastic analytical continuation, and stochastic optimization method, have been implemented in the ACFlow toolkit. \\
{\em Additional comments including restrictions and unusual features (approx. 50-250 words):} 
The ACFlow toolkit is written by pure Julia language. It is highly optimized and parallelized. It can be executed interactively in a Jupyter notebook environment. \\
\end{small}

\section{Introduction\label{sec:intro}}

It is well-known that quantum Monte Carlo (QMC) method is a powerful and exact numerical approach, and has been widely used in many research fields, such as nuclear physics~\cite{RevModPhys.87.1067}, condense matter physics~\cite{RevModPhys.73.33}, and many-body physics~\cite{RevModPhys.83.349}. In this paper, we just focus on the finite temperature QMC algorithms, which are used to solve the interacting lattice models or quantum impurity models~\cite{gubernatis_kawashima_werner_2016}. Generally speaking, the simulated results of QMC methods are some sorts of single-particle or two-particle correlation functions, which are usually defined on imaginary time axis ($\tau \equiv -it$) or Matsubara frequency axis ($i\omega_n$). Therefore, they can't be compared directly with the correspondingly experimental results, including but not limited to the electronic density of states $A(\omega)$, optical conductivity $\sigma(\omega)$, dynamical structure factor $S(\mathbf{q},\omega)$, and so on. It is necessary to convert the QMC simulated results from imaginary time axis or Matsubara frequency axis to real axis (i.e. $\tau \to \omega$ or $i\omega_n \to \omega$), which is the origin of the analytical continuation problem.

Let's concentrate on the following Fredholm integral equation of the first kind:
\begin{equation}
\label{eq:fredholm}
g(y) = \int K(y,x) f(x)~dx.
\end{equation}
Here, $K(y,x)$ is the known kernel function, $f(x)$ is the model function, and $g(y)$ denotes the raw data. Given $f(x)$, it is quite easy to get $g(y)$ via numerical integration. However, given $g(y)$, solving the Fredholm integral equation reversely to get $f(x)$ is not as easy as expected. There is no universal solution. Notice that the so-called analytical continuation problem can be reformulated in terms of the Fredholm integral equation. Thus, its objective is to seek a reasonable $f(x)$ to satisfy the above equation. The QMC simulated data $g(y)$ are noisy and the kernel function $K(y,x)$ is ill conditioned, which make analytical continuation of QMC simulated data a huge challenge. In order to solve this problem, peoples have developed numerous methods in the past decades. These methods include the least square fitting method, singular value decomposition~\cite{PhysRevB.82.165125,PhysRevLett.75.517}, Pad\'{e} approximation~\cite{Vidberg1977,PhysRevB.61.5147,PhysRevB.87.245135,PhysRevB.93.075104}, Tikhonov-Philips regularization method, maximum entropy method~\cite{PhysRevB.44.6011,JARRELL1996133}, stochastic analytical continuation~\cite{PhysRevB.57.10287,PhysRevE.81.056701}, stochastic optimization method~\cite{PhysRevB.62.6317,PhysRevB.95.014102}, sparse modelling method~\cite{PhysRevE.95.061302}, and machine learning method~\cite{PhysRevLett.124.056401,PhysRevB.98.245101,Arsenault_2017}, etc. However, each method has its pros and cons. None of these methods can override the others. The analytical continuation problem is still far away from being completely solved.

In recent years, quite a few analytical continuation codes have been released, including maxent~(by Mark Jarrell)~\cite{JARRELL1996133}, maxent (in ALPSCore)~\cite{LEVY2017149}, $\Omega$Maxent~\cite{PhysRevE.94.023303}, ana\_cont~\cite{KAUFMANN2023108519}, SOM~(in TRIQS)~\cite{KRIVENKO2019166,KRIVENKO2022108491}, Stoch~(in ALF)~\cite{ALF2022}, just to name a few. We note that the maximum entropy method has dominated this field for quite a long time. Thus most of these codes only implement the maximum entropy method~\cite{JARRELL1996133,LEVY2017149,PhysRevE.94.023303,KAUFMANN2023108519}. It is rather difficult to crosscheck the simulated results obtained by various analytical continuation methods. In addition, the features of the available codes are quite limited and hard to examine new algorithms. In order to fill in this gap, we would like to present a new open source toolkit, called ACFlow, for analytical continuation. This toolkit implements three primary analytical continuation methods, including the maximum entropy method, stochastic analytical continuation, and stochastic optimization method, within an united framework. It provides an easy-to-used library and application interface. Some diagnostic and analytical tools are also available. With ACFlow, the users can easily setup and execute analytical continuation calculations, and validate the obtained results. We believe that this toolkit will play a vital role in solving analytical continuation problems.   

The rest of this paper is organized as follows. In section~\ref{sec:problem}, background of the analytical continuation problem is introduced. In section~\ref{sec:methods}, basic principles and key ingredients of the three analytical continuation methods as implemented in the ACFlow toolkit are summarized. Section~\ref{sec:overview} gives a brief overview about ACFlow's main features and structures. Section~\ref{sec:usage} is the major part of this paper, it explains basic usage, input and output files of ACFlow. In order to demonstrate usefulness of this toolkit, four typical examples are illustrated in section~\ref{sec:examples}. Finally, section~\ref{sec:outlook} serves as a short conclusion.  

\section{Problem\label{sec:problem}}

\subsection{Finite temperature Green's functions}

Under the Wick's rotation $t \to i\tau$, the time evolution operator in the Heisenberg picture $e^{itH}$ will be replaced by $e^{-\tau H}$. Such a transformation will increase efficiency of QMC random walking and suppress numerical oscillation (when $t$ is large, the periodic oscillation of $e^{itH}$ is quite obvious). This is an important reason why most of the finite temperature QMC algorithms are formulated in imaginary time axis. The outputs of finite temperature QMC simulations are usually single-particle or two-particle correlation functions. For example, the single-particle Green's function $G(\tau)$ is defined as follows: 
\begin{equation}
G(\tau) = \langle \mathcal{T}_{\tau} d(\tau) d^{\dagger}(0) \rangle,
\end{equation}
where $\tau$ denotes imaginary time, $\mathcal{T}_{\tau}$ denotes time-ordered operator, and $d$ and $d^{\dagger}$ are annihilation and creation operators, respectively. The Matsubara Green's function $G(i\omega_n)$ can be measured by QMC simulations or constructed from $G(\tau)$ via direct Fourier transformation:
\begin{equation}
G(i\omega_n) = \int^{\beta}_0 d\tau~e^{-i\omega_n \tau} G(\tau),
\end{equation}
\begin{equation}
G(\tau) = \frac{1}{\beta} \sum_n e^{i\omega_n \tau} G(i\omega_n).
\end{equation}
Here, $\beta$ means the inverse temperature ($\beta \equiv 1/T$) and $\omega_n$ is the Matsubara frequency. Note that $\omega_n$ is equal to $(2n + 1) \pi / \beta$ for fermions and $2n\pi/ \beta$ for bosons ($n$ is an integer).

\subsection{Spectral density}

Clearly, neither $G(\tau)$ nor $G(i\omega_n)$ can be observed experimentally. We have to extract dynamical response function, i.e., the spectral density $A(\omega)$, from them. $A(\omega)$ is indeed an observable quantity. It is related to $G(\tau)$ via the following Laplace transformation:  
\begin{equation}
\label{eq:spectral_density_1}
G(\tau) = \int^{+\infty}_{-\infty} d\omega \frac{e^{-\tau\omega}}{1 \pm e^{-\beta\omega}} A(\omega),
\end{equation} 
where +(-) in the denominator is for fermionic (bosonic) system. $G(i\omega_n)$ and $A(\omega)$ manifest similar relation:
\begin{equation}
\label{eq:spectral_density_2}
G(i\omega_n) = \int^{+\infty}_{-\infty} d\omega \frac{A(\omega)}{i\omega_n - \omega}.
\end{equation}
It is obvious that Eq.~(\ref{eq:spectral_density_1}) and Eq.~(\ref{eq:spectral_density_2}) are indeed two special forms of the Fredholm integral equation of the first kind [see Eq.~(\ref{eq:fredholm})]. So, the central problem of analytical continuation is to search optimal $A(\omega)$ for given $G(\tau)$ or $G(i\omega_n)$.

Sometimes the spectral density $A(\omega)$ is called spectral function in the references. It is tied to the imaginary part of real frequency Green's function $G(\omega)$:
\begin{equation}
\label{eq:ImG}
A(\omega) = -\frac{1}{\pi} \rm{Im}G(\omega).
\end{equation}
From Im$G(\omega)$, Re$G(\omega)$ could be calculated via the Kramers-Kronig transformation:
\begin{equation}
\label{eq:kk}
\mathrm{Re} G(\omega) = \frac{1}{\pi} \mathcal{P}
  \int_{-\infty}^{\infty} d\omega'~
  \frac{\mathrm{Im} G(\omega')}{\omega'-\omega},
\end{equation}
where $\mathcal{P}$ means Cauchy principal value. Besides Eq.~(\ref{eq:spectral_density_1}) and Eq.~(\ref{eq:spectral_density_2}), $A(\omega)$ has to obey some additional constraints or sum-rules. For fermionic systems, the spectral functions must be positive:
\begin{equation}
A(\omega) \ge 0.
\end{equation}
While for bosonic systems, the constraint becomes:
\begin{equation}
\text{sign}(\omega) A(\omega) \ge 0.
\end{equation}
In addition, the spectral function $A(\omega)$ is always bounded,
\begin{equation}
\int^{+\infty}_{-\infty} d\omega~A(\omega) < \infty.
\end{equation}
It can be utilized to normalize the resulting spectral function.

\subsection{Kernel functions}

Eq.~(\ref{eq:spectral_density_1}) and Eq.~(\ref{eq:spectral_density_2}) can be reformulated as follows:
\begin{equation}
\label{eq:kernel_t}
G(\tau) = \int^{+\infty}_{-\infty} d\omega~K(\tau,\omega) A(\omega),
\end{equation}
and
\begin{equation}
\label{eq:kernel_w}
G(i\omega_n) = \int^{+\infty}_{-\infty} d\omega~K(\omega_n,\omega) A(\omega),
\end{equation}
where $K(\tau,\omega)$ and $K(\omega_n, \omega)$ are the so-called kernel functions. Their definitions are as follows:
\begin{equation}
\label{eq:ktau}
K(\tau,\omega) = \frac{e^{-\tau\omega}}{1 \pm e^{-\beta\omega}},
\end{equation}
and
\begin{equation}
\label{eq:komega}
K(\omega_n,\omega) = \frac{1}{i\omega_n - \omega},
\end{equation}
where +(-) in the denominator of Eq.~(\ref{eq:ktau}) stands for fermions (bosons).

As mentioned above, the kernel function is quite strange. The values of $K(\tau,\omega)$ could change by tens of orders of magnitude. Especially, at large positive and negative frequencies, $K(\tau,\omega)$ is exponentially small. It implies that at large $|\omega|$ the features of $A(\omega)$ are sensitive to the fine structures of $G(\tau)$. However, the data of $G(\tau)$ provided by QMC simulations are always fluctuant and noisy~\cite{PhysRevB.84.075145}. Tiny deviations in $G(\tau)$ from its expected values can lead to enormous changes in $A(\omega)$. Thus, analytical continuation is often characterized as an ill-posed problem~\cite{JARRELL1996133}.

In principle, for incomplete and noise $G(\tau)$ or $G(i\omega_n)$, the number of spectral functions $A(\omega)$ that satisfy Eq.~(\ref{eq:kernel_t}) and Eq.~(\ref{eq:kernel_w}) is infinite. So the question becomes which $A(\omega)$ should be chosen. Now there are two different strategies to solve this problem. The first one is to choose the most likely $A(\omega)$. The second one is to evaluate the average of all the candidate spectral functions. In next section, we will introduce three primary analytical continuation methods that follow the two strategies and have been implemented in the ACFlow toolkit. For the sake of simplicity, we will concentrate on analytical continuation of imaginary time Green's functions in main text. 
 
\section{Methods\label{sec:methods}}

\subsection{Maximum entropy method}

Perhaps the maximum entropy method is the most frequently used approach for analytical continuation problems in the last decades~\cite{PhysRevB.44.6011,JARRELL1996133} because of its high computational efficiency. Next, we will discuss the basic principle and several variants of it.

\subsubsection{Bayesian inference}

Bayes's theorem is the cornerstone of the maximum entropy method. Given two events $a$ and $b$, Bayes's theorem says:
\begin{equation}
P[a|b]P[b] = P[b|a]P[a],
\end{equation}
where $P[a]$ is the probability of event $a$, $P[a|b]$ is the conditional probability of event $a$ with given event $b$. In the scenario of analytical continuation problem, $\bar{G}(\tau)$ and $A(\omega)$ are treated as two events, where $\bar{G}(\tau)$ denotes the measured value of $G(\tau)$. So the best solution for $A(\omega)$ is of course the one that maximizes $P[A|\bar{G}]$, which is called the posterior probability. According to the Bayes's theorem, we get
\begin{equation}
P[A|\bar{G}] = \frac{P[\bar{G}|A]P[A]}{P[\bar{G}]},
\end{equation}
where $P[\bar{G}|A]$ is the likelihood function, $P[A]$ is the prior probability, and $P[\bar{G}]$ is the evidence. Since the evidence is a normalization constant depending on the prior probability and the likelihood function only, it is ignored in the following discussions. Thus,
\begin{equation}
P[A|\bar{G}] \propto P[\bar{G}|A]P[A].
\end{equation}

\subsubsection{Posterior probability}

In the maximum entropy method, the likelihood function $P[\bar{G}|A]$ is assumed to be in direct proportion to $e^{-\chi^2/2}$. Here, $\chi^2$ is named as goodness-of-fit function. It measures the distance between $\bar{G}(\tau)$ and reconstructed imaginary time Green's function $\tilde{G}(\tau)$:   
\begin{equation}
\label{eq:chi2}
\chi^2 = \sum^{L}_{i = 1} \left[\frac{\bar{G}_i(\tau) - \tilde{G}_i(\tau)}{\sigma_i}\right]^2,
\end{equation}
\begin{equation}
\tilde{G}_i = \sum_j K_{ij} A_j.
\end{equation}
Here, $L$ is number of imaginary time points, $\sigma$ denotes the error bar (standard deviation) of $\bar{G}(\tau)$. $K_{ij}$ and $A_j$ are discrete kernel and spectral functions, respectively. On the other hand, the prior probability $P[A]$ is supposed to be in direct proportion to $e^{\alpha S}$, where $\alpha$ is a regulation parameter and $S$ means entropy. Sometimes $S$ is also known as the Kullback-Leibler distance. Its formula is as follows:
\begin{equation}
S= \int d\omega \left(A(\omega) - m(\omega) - A(\omega)\log\left[\frac{A(\omega)}{m(\omega)}\right]\right),
\end{equation}
where $m(\omega)$ is the default model function.

According to the Bayes's theorem, the posterior probability $P[A|\bar{G}] \propto e^{Q}$ and
\begin{equation}
Q = \alpha S - \frac{\chi^2}{2}.
\end{equation}

\subsubsection{Algorithms of maximum entropy method}

Now the original analytical continuation problem becomes how to figure out the optimal $A(\omega)$ that maximizes $Q$. In other words, we have to solve the following equation:
\begin{equation}
\label{eq:maxent}
\frac{\partial Q}{\partial A} \bigg|_{A = \hat{A}} = 0,
\end{equation}
where $\hat{A}(\omega)$ is the optimal $A(\omega)$. Eq.~(\ref{eq:maxent}) can be easily solved by using standard Newton method. However, the obtained $\hat{A}(\omega)$ is $\alpha$-dependent. That is to say, for a given $\alpha$, there is always a $\hat{A}(\omega)$ that satisfies Eq.~(\ref{eq:maxent}). So, new problem arises because we have to figure out a way to construct the final spectral function from these $\alpha$-resolved $\hat{A}(\omega)$. Now there exist four algorithms, namely ``historic'', ``classic'', ``bryan'', and ``$\chi^2$kink''. Next we will introduce them one by one.

\emph{Historic algorithm}. The historic algorithm is quite simple. The $\alpha$ parameter will be adjusted iteratively to meet the following criterion:
\begin{equation}
\chi^2 = N,
\end{equation}  
where $N$ is the number of mesh points for spectral density $A(\omega)$.

\emph{Classic algorithm}. The basic equation for the classic algorithm reads:
\begin{equation}
\label{eq:classic}
-2 \alpha S(A_{\alpha}) = \text{Tr} 
\left[
\frac{\Lambda(A_{\alpha})}{\alpha I + \Lambda(A_{\alpha})}
\right],
\end{equation}
where $I$ is an identity matrix. The elements of $\Lambda$ matrix are calculated as follows:
\begin{equation}
\Lambda_{ij} = \sqrt{A_i} \left(\sum_{kl} K_{ki} [C^{-1}]_{kl} K_{lj}\right) \sqrt{A_j}, 
\end{equation}
where $C$ is the covariance matrix. Eq.~(\ref{eq:classic}) will be iteratively solved until the optimal $\alpha$ and $\hat{A}(\omega)$ are determined.  

\emph{Bryan algorithm}. In both historic and classic algorithms, the spectral function $\hat{A}(\omega)$ is always related to an optimal $\alpha$ parameter. However, the spirit of the bryan algorithm~\cite{Bryan1990} is completely different. It tries to generate a series of $\alpha$ parameters and yield the corresponding $A_{\alpha}(\omega)$. Then the final spectral function $A(\omega)$ is obtained by evaluating the following integration:
\begin{equation}
\overline{A(\omega)} = \int d\alpha~A_{\alpha}(\omega) P[\alpha | \bar{G}].
\end{equation}

\emph{$\chi^{2}$kink algorithm}. This algorithm was proposed by Bergeron and Tremblay~\cite{PhysRevE.94.023303} recently. The first step is to generate a series of $\alpha$ parameters, and evaluate the corresponding spectral functions $A_{\alpha}(\omega)$ and the goodness-of-fit functions $\chi^{2}[A_{\alpha}]$. Then we plot $\log_{10}(\chi^{2})$ as a function of $\log_{10}(\alpha)$. Usually this plot is split into three different regions: (1)~Default model region. In the limit of $\alpha \to \infty$, $\chi^{2}$ goes to a constant high value. It means that the likelihood function $e^{-\chi^2/2}$ has negligible weight, such that the prior probability $e^{\alpha S}$ becomes dominant and minimizes $Q[A]$. At that time, the calculated $A(\omega)$ resembles the default model function $m(\omega)$. (2)~Noise-fitting region. In the limit of $\alpha \to 0$, $\chi^2$ is relatively flat and approaches its global minimum. In this region, the minimization algorithm tends to fit the noise in $G(\tau)$. (3)~Information-fitting region. $\alpha S$ is comparable with $\chi^2/2$, so that $\chi^{2}$ is strongly dependent on $\alpha$. Bergeron \emph{et al.} suggested that the optimal $\alpha$ parameter situates in the crossover between noise-fitting region and information-fitting region~\cite{PhysRevE.94.023303}. So the second derivative of $\chi^{2}$ with respect to $\alpha$ is calculated, and the maximum value in the resulting curve indicates the optimal value of $\alpha$. Quite recently, Kaufmann and Held proposed a more numerically stable and flexible approach to compute the optimal $\alpha$~\cite{KAUFMANN2023108519}. They use the following empirical function to fit dataset $\{\log_{10} (\alpha), \log_{10} (\chi^{2})\}$:
\begin{equation}
\phi(x;a,b,c,d) = a + \frac{b}{1 + e^{-d(x-c)}},
\end{equation}
where $a$, $b$, $c$, and $d$ are fitting parameters. Then the optimal $\alpha$ is approximated by $10^{c-f/d}$, where $f$ is a numerical constant (Its favorite value lies in $[2,2.5]$).

\subsection{Stochastic analytical continuation}

In principle, for given Green's function $G$, there exists infinitely many spectral densities $A(\omega)$ that can be used to reconstruct $G$ via Eq.~(\ref{eq:kernel_t}) and Eq.~(\ref{eq:kernel_w}). The maximum entropy method tries to pick up the most likely spectral function which maximizes $P[A|\bar{G}]$ (It actually maximizes $Q$)~\cite{PhysRevB.44.6011,JARRELL1996133}. Here, we would like to introduce an alternative approach, namely the stochastic analytical continuation~\cite{PhysRevB.57.10287,PhysRevE.81.056701,beach,PhysRevX.7.041072,PhysRevE.94.063308,PhysRevLett.118.147207,PhysRevLett.63.1523,Shao:2022yez}. It is argued that the weights for all the possible spectral densities are the same if they can give rise to the same $\chi^2$. At first, a sequence of spectral densities will be generated by stochastic method. Then an unbiased thermal average of all possible spectra, Boltzmann weighted according to goodness-of-fit function $\chi^{2}$, produces an average spectrum. Thus sometimes the method was named as average spectrum method or stochastic sampling method in the references~\cite{PhysRevB.101.085111,PhysRevB.102.035114,PhysRevB.76.035115,PhysRevB.78.174429}. There are several variants for the stochastic analytical continuation. Next we will introduce two representative algorithms as proposed by A. W. Sandvik~\cite{PhysRevB.57.10287} and K. S. D. Beach~\cite{beach}, respectively.

\subsubsection{Sandvik's algorithm}

\begin{figure}[ht]
\centering
\includegraphics[width=0.7\textwidth]{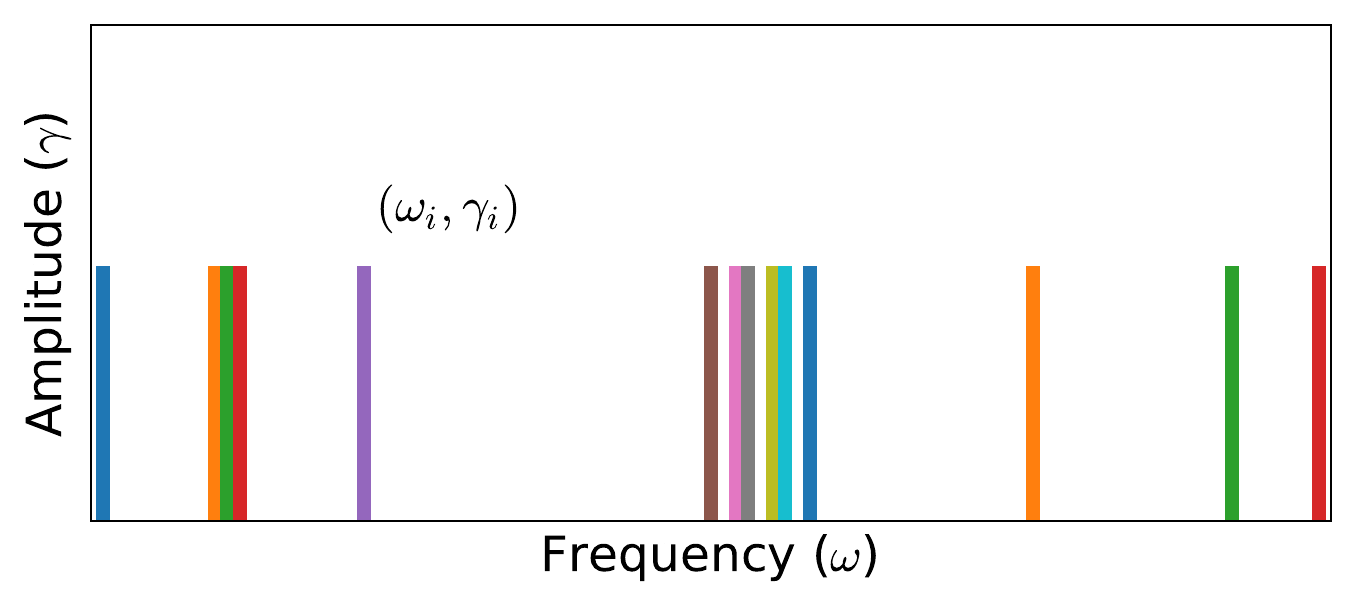}
\caption{Typical Monte Carlo field configurations for the stochastic analytical continuation (A. W. Sandvik's algorithm)~\cite{PhysRevB.57.10287}. Here, the $\delta$ functions reside at unrestricted frequencies $\{\omega_i\}$, but their amplitudes $\{\gamma_i\}$ are equal and fixed. Note that different parameterizations are also possible~\cite{Shao:2022yez}. \label{fig:san}}
\end{figure}

It was early on realized that a different way to achieve a smooth spectrum is to average over many solutions with reasonable $\chi^2$ values~\cite{PhysRevLett.63.1523}. Several years later, A. W. Sandvik introduced the stochastic analytical continuation in a slightly different form~\cite{PhysRevB.57.10287}. He suggested that the spectral function $A(\omega)$ can be parameterized using $N$ $\delta$ functions (Please see Figure~\ref{fig:san} for a schematic diagram):
\begin{equation}
A(\omega) = \sum^{N}_{i = 1} \gamma_i \delta(\omega - \omega_i),
\end{equation}
where $\gamma_i$ and $\omega_i$ denote the amplitude and position of the $i$-th $\delta$ function, respectively. Next, the Metropolis important sampling algorithm is employed to sample the configuration space $\mathcal{C} = \{\omega_i, \gamma_i\}$. In practice, there are two elementary Monte Carlo updates. One is to change the amplitudes of a pair of $\delta$ functions under the constraint $\sum_i \gamma_i = 1$. Another one is to shift position of a randomly chosen $\delta$ function. Of course, block or global updates can be implemented to improve ergodicity and sampling efficiency~\cite{PhysRevB.101.085111,PhysRevB.102.035114}. 

The transition probability of Monte Carlo updates reads:
\begin{equation}
\label{eq:trans_san}
p(\mathcal{C} \to \mathcal{C}') = \exp\left(-\frac{\Delta\chi^2}{2\Theta}\right),
\end{equation}
where the goodness-of-fit function $\chi^2$ can be evaluated by Eq.~(\ref{eq:chi2}), $\Theta$ is a regulation parameter which is similar to the $\alpha$ parameter appeared in the maximum entropy method. Well, the remaining problem is how to fix $\Theta$. Sandvik suggested to measure the following entropic term for a series of $\Theta$:  
\begin{equation}
S(\Theta) = - \sum^{N}_{i = 1} \gamma_i \log(\gamma_i) K(0,\omega_i),
\end{equation}
where $K$ is the kernel function as defined above~\cite{PhysRevB.57.10287}. Then make a plot of $S$ with respect to $\log(\Theta^{-1})$. Overall, when $\Theta$ is large, $S$ exhibits large fluctuations. When $\Theta$ is small, $S$ will approach its global minimum steadily. A sharp drop in $S$ before the approach to a constant value has been observed, and there is a local maximum at some $\Theta = \hat{\Theta}$ preceding the drop. Thus, Sandvik postulated that $\hat{\Theta}$ was the optimum value at which to accumulate and average the spectral function. Sylju\aa{}sen \emph{et al.}~\cite{PhysRevB.78.174429} suggested that let $\Theta = 1$. Fuchs \emph{et al.} tried to fix $\Theta$ by using Bayesian inference. Such that their approach was named as stochastic analytical inference~\cite{PhysRevE.81.056701}. Very recently, Shao and Sandvik \emph{et al.} proposed a smart method to determine the optimal value of $\Theta$~\cite{PhysRevLett.118.147207,PhysRevX.7.041072}. $\Theta$ is adjusted so that
\begin{equation}
\langle \chi^2(\Theta) \rangle \approx \chi^{2}_{\text{min}} + c \sqrt{2\chi^2_{\text{min}}},
\end{equation}
where $c$ is a constant of order 1, $\chi^{2}_{\text{min}}$ is the minimum value of $\chi^{2}$ at given $\Theta$. Note that $\chi^{2}_{\text{min}}$ can be obtained in a simulated annealing process~\cite{SA1983} to very low $\Theta$.

\subsubsection{Beach's algorithm}

\begin{figure}[ht]
\centering
\includegraphics[width=0.7\textwidth]{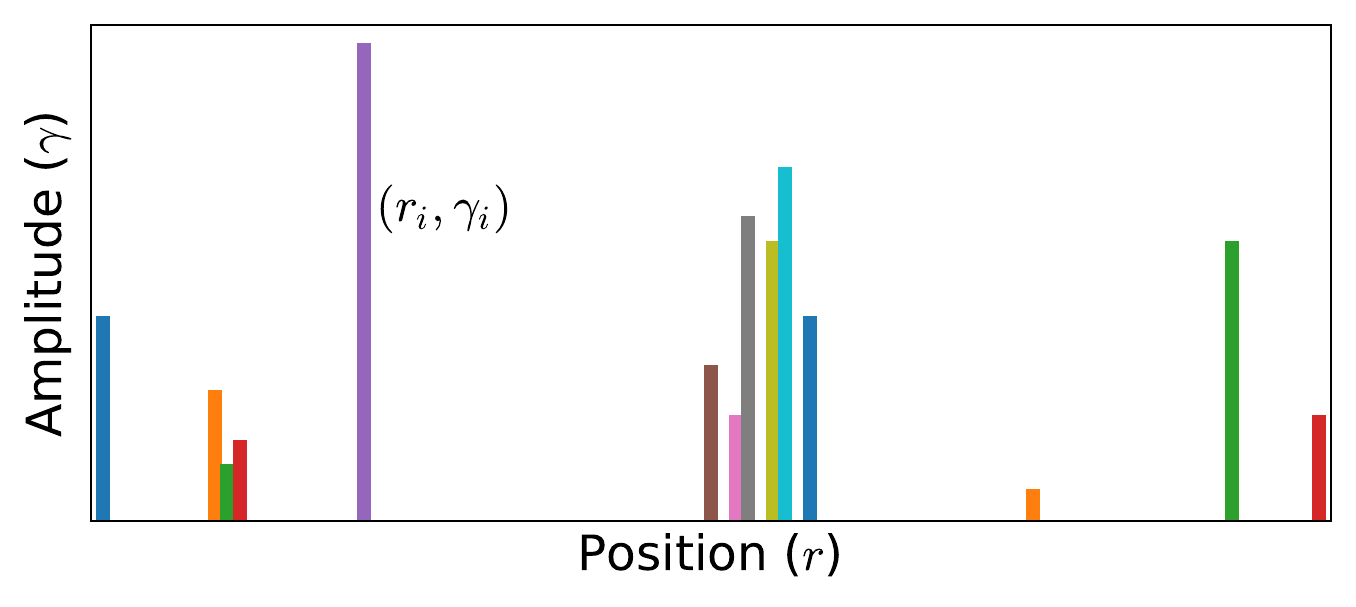}
\caption{Typical Monte Carlo field configurations for the stochastic analytical continuation (K. S. D. Beach's algorithm)~\cite{beach}. Note that the amplitudes $\{\gamma_i\}$ of all the $\delta$ functions are not identical. Both amplitudes $\{\gamma_i\}$ and positions $\{r_i\}$ ($0.0 < r_i < 1.0$) can be sampled by using Monte Carlo method. \label{fig:sac}}
\end{figure}

K. S. D. Beach proposed another variant of stochastic analytical continuation in 2004~\cite{beach}. In his proposal, the analytical continuation problem is mapped into a system of interacting classic fields at first. Then the classic field is sampled using Monte Carlo method to obtain the final solution. He concluded that the maximum entropy method is simply the mean field limit of the stochastic analytical continuation. Next, this algorithm will be explained concisely.     

\emph{Classic fields}. Recalled that the goodness-of-fit functional $\chi^{2}[A]$ measures how closely the Green's function generated from $A(\omega)$ matches the raw input data. Its expression is rewritten as follows: 
\begin{equation}
\chi^{2}[A] = \int^{\beta}_{0} \frac{1}{\sigma(\tau)^2} 
\left|\int d\omega~K(\tau,\omega) A(\omega) - \bar{G}(\tau)\right|^2 d\tau.
\end{equation}
At first, a new variable $x$ is introduced. The relation between $x$ and $\omega$ is:  
\begin{equation}
x = \phi(\omega) = \int^{\omega}_{-\infty} d\omega'~m(\omega'),
\end{equation}
where $m(\omega)$ denotes the default model function. Clearly, the $\phi(\omega)$ function defines a smooth mapping from $\mathbf{R} \to [0,1]$. Since $\omega = \phi^{-1}(x)$, a dimensionless classic field $n(x)$ is created:
\begin{equation}
n(x) = \frac{A(\phi^{-1}(x))}{m(\phi^{-1}(x))}.
\end{equation}
It is easy to prove that both $n(x)$ and $A(\omega)$ obey similar normalization condition: 
\begin{equation}
\int d\omega~A(\omega) = \int^{1}_0 dx~n(x) = 1.
\end{equation}
Next, in analogy with the goodness-of-fit functional $\chi^2[A]$, the Hamiltonian for the system of classic field $\{n(x)\}$ can be defined as follows:
\begin{equation}
\label{eq:hamil}
H[n(x)] = \int^{\beta}_0 \frac{d\tau}{\sigma(\tau)^2}
\left|
\int^{1}_0 dx~K(\tau,x) n(x) - \bar{G}(\tau)
\right|.
\end{equation}
Supposing $\alpha$ is an inverse temperature of the system, then the partition function $Z$ is:
\begin{equation}
Z = \int \mathcal{D}n~e^{-\alpha H[n]},
\end{equation}
where
\begin{equation}
\int \mathcal{D}n = 
\int^{\infty}_0 \left[\prod_x dn(x)\right]
\delta\left(\int^{1}_0 dx~n(x) - 1\right).
\end{equation}
The thermally averaged value of the classic field is:
\begin{equation}
\langle n(x) \rangle = \frac{1}{Z} \int \mathcal{D}n~n(x) e^{-\alpha H[n]}.
\end{equation}
Finally, according to the definition of the classic field, the averaged spectral density $\langle A(\omega) \rangle$ can be expressed as:
\begin{equation}
\langle A(\omega) \rangle = \langle n(\phi(\omega)) \rangle m(\omega).
\end{equation}
So, by introducing the classic field $\{n(x)\}$, the analytical continuation problem is converted into a statistical sampling of the classic field, which is easily solved by using Monte Carlo method.   

\emph{Monte Carlo sampling}. Next we clarify how to sample the classic field. Similar to Sandvik's algorithm~\cite{PhysRevB.57.10287,Shao:2022yez}, $n(x)$ is parameterized as a superposition of many $\delta$ functions (see Figure~\ref{fig:sac} for a schematic diagram): 
\begin{equation}
n_{\mathcal{C}} (x) = \sum_i \gamma_i \delta(x - r_i),
\end{equation}
where $\gamma_i$ and $r_i$ denote amplitude (weight) and position of the $i$-th $\delta$ function, respectively. And $\mathcal{C}$ means a configuration space formed by a set of $r_i$ and $\gamma_i$,
\begin{equation}
\mathcal{C} = \{r_i, \gamma_i\}.
\end{equation}
Note that $\gamma_i$ and $r_i$ satisfy the following constraints:
\begin{equation}
\forall i,~\gamma_i > 0,~\sum_i \gamma_i = 1,~ 0 \le r_i \le 1.
\end{equation}
Supposed that there is a transition from $\mathcal{C}$ to $\mathcal{C}'$ ($\{r_i, \gamma_i\} \to \{r'_i, \gamma'_i\}$): 
\begin{equation}
r_i \to r'_i = 
r_i + \sum_{\lambda \in \Lambda} \delta_{i\lambda} \Delta r_{\lambda},
\end{equation}
\begin{equation}
\gamma_i \to \gamma'_i =
\gamma_i + \sum_{\lambda \in \Lambda} \delta_{i\lambda} \Delta \gamma_{\lambda},
\end{equation}
where $\Lambda$ means a subset of the $\delta$ functions, then the Hamiltonian of the system is changed from $H_{\mathcal{C}}$ to $H_{\mathcal{C}'}$. According to Eq.~(\ref{eq:hamil}), $H_\mathcal{C}$, $H_{\mathcal{C}'}$, and their difference $\Delta H$ can be calculated by:
\begin{equation}
H_{\mathcal{C}} = \int^{\beta}_0 d\tau~h_{\mathcal{C}}(\tau)^2,
\end{equation}
\begin{equation}
H_{\mathcal{C}'} = \int^{\beta}_0 d\tau 
\left[h_{\mathcal{C}}(\tau) + \Delta h(\tau)\right]^2,
\end{equation}
\begin{equation}
\Delta H = H_{\mathcal{C}'} - H_{\mathcal{C}} = 
\int^{\beta}_0 d\tau~\Delta h(\tau) 
[2h_{\mathcal{C}}(\tau) + \Delta h(\tau)].
\end{equation}
Here,
\begin{equation}
h(\tau) = \frac{1}{\sigma(\tau)} \left[\int^1_0 dx~K(\tau, x)n(x) - \bar{G}(\tau)\right],
\end{equation}
and
\begin{equation}
\Delta h(\tau) = \frac{1}{\sigma(\tau)}
\sum_{\lambda \in \Lambda}
\left[
\gamma'_{\lambda} K(\tau,r'_{\lambda}) - \gamma_{\lambda} K(\tau,r_{\lambda})
\right].
\end{equation}
Finally, the transition probability from $\mathcal{C}$ to $\mathcal{C}'$ reads
\begin{equation}
\label{eq:trans_sac}
p(C \to C') = \exp(-\alpha \Delta H).
\end{equation}

\emph{Parallel tempering}. The parallel tempering trick~\cite{Marinari:1996dh} is adopted to improve the Monte Carlo algorithm as described above. It is possible to proceed multiple simulations simultaneously for a sequence of inverse temperature parameters $\{\alpha_1, \alpha_2, \cdots, \alpha_N \}$. The ratio for two adjacent $\alpha$ parameters is a constant: $\alpha_{p+1} / \alpha_p = R$. Note that the field configurations in all simulations evolve in parallel but not independently. We can swap the field configurations between two adjacent layers. Of course, the detailed balance is always preserved, and each simulation will eventually settle into thermal equilibrium at given $\alpha$. The transition probability of such a global Monte Carlo update is:
\begin{equation}
p(\mathcal{C} \to \mathcal{C}') = \exp[(\alpha_p - \alpha_q)(H_{p} - H_{q})],
\end{equation}
where $p$ and $q$ are layer indices, and $p = q \pm 1$. Parallel tempering eliminates the need for an initial annealing stage. Another advantage of parallel tempering is that it yields a complete temperature profile of all the important thermodynamic variables (such as specific heat and internal energy), which can be used to estimate the critical $\alpha$ and the final spectral function $\langle A(\omega) \rangle$.   

\emph{Critical inverse temperature}. Clearly, $\langle n(x) \rangle$ strongly depends on the inverse temperature $\alpha$. How to use these $\alpha$-dependent $\langle n(x) \rangle$ to construct the final spectral function? Beach suggested a novel method~\cite{beach}. During parallel tempering process, the internal energy of the system is also measured in addition to $\langle n(x) \rangle$: 
\begin{equation}
U(\alpha_p) = \langle H [n] \rangle_{\alpha_p}.
\end{equation}
Let us plot $\log_{10}[U(\alpha)]$ as a function of $\log_{10} (\alpha)$. We find that $\log_{10}[U(\alpha)]$ drops quickly at first when $\log_{10} (\alpha)$ increases, and then it approaches to a constant value slowly. The knee in $\log_{10}[U(\alpha)]$ function, occurring in the vicinity of $\alpha = \alpha^*$ (the corresponding layer index $p = p^{*}$), signals a jump in specific heat (a thermodynamic phase transition). Then the averaged spectral function is constructed by:
\begin{equation}
\langle \langle n(x) \rangle \rangle =
\frac{\sum^{N-1}_{p = p*} [U(\alpha_p) - U(\alpha_{p+1})] \langle n(x) \rangle_{\alpha_p}}
{U(\alpha_{p*}) - U(\alpha_N)},
\end{equation}
where $N$ is the total number of $\alpha$, and $\alpha_{p*}$ ($\equiv \alpha^{*}$) is the critical inverse temperature.

\emph{Likelihood function}. Neither of the Sandvik's and Beach's algorithms needs extra entropic term to regulate the spectral densities~\cite{PhysRevB.57.10287,beach}. All the stochastically generated spectra are treated on the same footing. Thus, the calculated spectral function retains more subtle structures than that obtained by the maximum entropy method. Actually, in the stochastic analytical continuation, 
\begin{equation}
\langle A \rangle = \int \mathcal{D} A~P[A|\bar{G}] A.
\end{equation}
The weight of the candidate spectral function $A$ is given by the likelihood function $P[A|\bar{G}]$. Eq.~(\ref{eq:trans_san}) and Eq.~(\ref{eq:trans_sac}) can be viewed as likelihood functions in the stochastic analytical continuation.

\subsection{Stochastic optimization method}

\begin{figure}[th]
\centering
\includegraphics[width=0.7\textwidth]{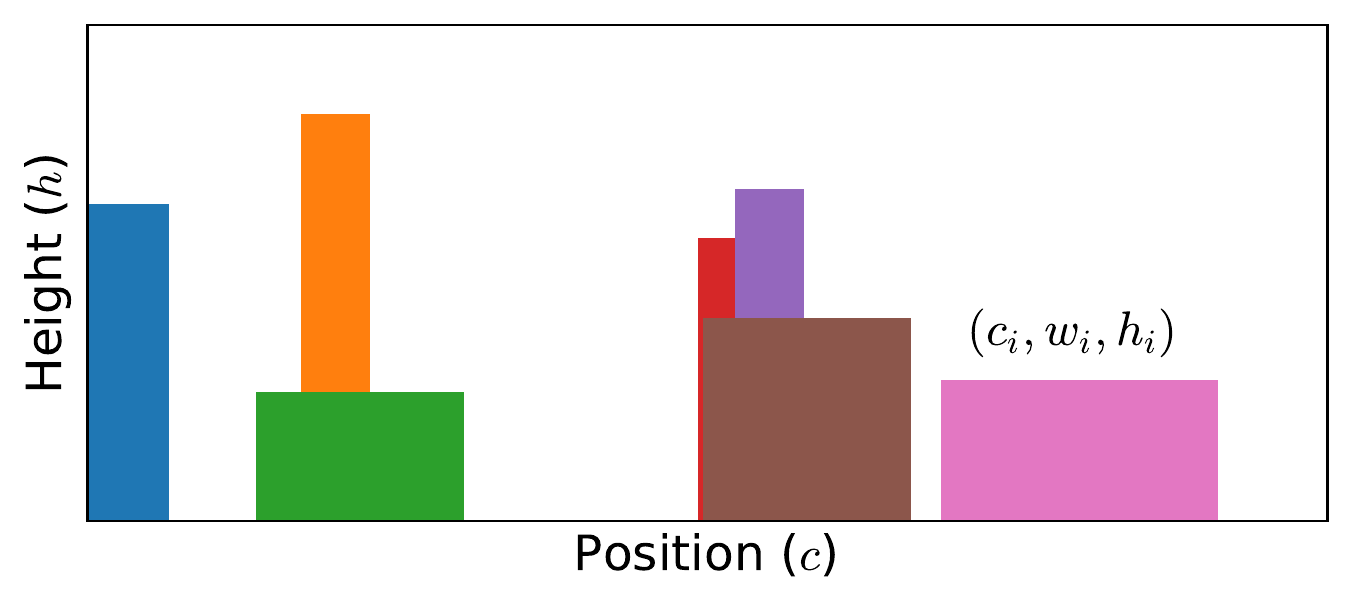}
\caption{Typical Monte Carlo field configurations for the stochastic optimization method~\cite{PhysRevB.62.6317}. The spectral function is parameterized by multiple rectangle functions. Here, $c_i$, $w_i$, and $h_i$ denote the center, width, and height of the $i$-th rectangle, respectively. \label{fig:som}}
\end{figure}

A. O. Mishchenko \emph{et al.}~\cite{PhysRevB.62.6317} proposed the stochastic optimization method. Though it looks like the stochastic analytical continuation~\cite{PhysRevB.57.10287,beach}, their differences are quite apparent. The stochastic optimization method does not need any likelihood function or Boltzmann distribution to weight the candidate spectral functions. It generates a lot of spectral functions through Monte Carlo samplings. For each candidate spectral function, the deviation $D$ between the reconstructed Green's function $\tilde{G}$ and original Green's function $\bar{G}$ is measured. Those spectral functions with small deviations $D$ are selected and averaged. Such that the desired spectral function is obtained.   

\emph{Deviation function}. In the stochastic optimization method, the deviation between reconstructed data $\tilde{G}$ and input data $\bar{G}$ is described by:
\begin{equation}
D[A] = \sum^{M}_{m=1} |\Delta(m)|,
\end{equation}
where $M$ is the number of input data, and $\Delta(m)$ is the deviation function,
\begin{equation}
\Delta(m) = \frac{\bar{G}(m) - \tilde{G}(m)}{S(m)}.
\end{equation}
Here, $S(m) = |G(m)|^{d}$ (where $0 \le d \le 1$). Recently, Krivenko \emph{et al.} suggested that it would be better to use the goodness-of-fit functional $\chi^2[A]$ to replace $D[A]$~\cite{KRIVENKO2019166,KRIVENKO2022108491}.  

\emph{Spectral density}. The stochastic optimization method will try to accumulate the candidate spectral functions that manifest small $D[A]$. Supposed the Monte Carlo simulations are repeated for $L$ times. For the $i$-th Monte Carlo simulation, the spectral density $A_i(\omega)$ and deviation $D[A_i]$ are recorded. The minimum value of deviation is $\min\{D[A_i]\}$. Thus, the final spectral density reads: 
\begin{equation}
A(\omega) = \frac{1}{L_{\text{good}}} \sum^{L}_{i = 1}
\theta(\alpha_{\text{good}} \min\{D[A_i]\} - D[A_i]) A_{i}(\omega).
\end{equation}
Here, $\theta(x)$ is the Heaviside step function, and $\alpha_{\text{good}}$ is a adjustable parameter. $L_{\text{good}}$ denotes the number of ``good'' spectral functions:
\begin{equation}
L_{\text{good}} = \sum^{L}_{i = 1} 
\theta(\alpha_{\text{good}} \min\{D[A_i]\} - D[A_i]).
\end{equation}
That is to say, only those spectral functions who satisfy the following condition will be selected:
\begin{equation}
D[A_i] \le \alpha_{\text{good}} \min\{D[A_i]\}.
\end{equation}
Clearly, the larger $\alpha_{\text{good}}$ is, the more spectral functions are included. It is usually set to 2.  

\emph{Rectangle representation}. Similar to the stochastic analytical continuation~\cite{PhysRevB.57.10287,beach}, the stochastic optimization method usually employs a few rectangle functions to parameterize the spectral function:
\begin{equation}
A(\omega) = \sum_i R_{\{c_i, w_i, h_i\}} (\omega),
\end{equation}
where $i$ is the index of rectangle function. The definition of rectangle function $R_{\{c_i, w_i, h_i\}} (\omega)$ reads:
\begin{equation}
R_{\{c_i, w_i, h_i\}} (\omega) = 
h_i 
\theta[\omega - (c_i - w_i/2)]
\theta[(c_i + w_i/2) - \omega],
\end{equation}
where $c_i$, $w_i$, $h_i$ denote the center, width, and height of the $i$-th rectangle, respectively. Pay attention to that the area of all rectangles must be normalized to 1:
\begin{equation}
\sum_i h_i w_i = 1.
\end{equation}

\emph{Monte Carlo sampling}. The parameters of all rectangle functions create a configuration space:
\begin{equation}
\mathcal{C} = \{c_i, w_i, h_i\}.
\end{equation}
Then the Metropolis algorithm is utilized to sample this configuration space. Mishchenko \emph{et al.} introduces seven Monte Carlo updates~\cite{PhysRevB.62.6317,KRIVENKO2019166}, including: (a) Insert a new rectangle, change width and height of another rectangle; (b) Remove an existing rectangle, change width and height of another rectangle; (c) Shift position of any rectangles; (d) Change widths of any two rectangles;  (e) Change heights of any two rectangles; (f) Split a rectangle into two new rectangles; (g) Merge two adjacent rectangles into a new rectangle. The transition probability of these Monte Carlo updates is:
\begin{equation}
p(\mathcal{C} \to \mathcal{C}') = \left(\frac{D[A_{\mathcal{C}}]}{D[A_{\mathcal{C}'}]}\right)^{1+d}
\end{equation}

As compared to the maximum entropy method~\cite{PhysRevB.44.6011,JARRELL1996133}, the likelihood function, entropic term, and model function are absent in the stochastic optimization method. As compared to the stochastic analytical continuation~\cite{PhysRevB.57.10287,beach}, there are no adjustable parameters, such as $\Theta$ in Sandvik's algorithm and $\alpha$ in Beach's algorithm. Thus, the simulated results of the stochastic optimization method are less affected by artificial parameters.       

\section{Overview\label{sec:overview}}

\subsection{Major features}

Now the ACFlow toolkit supports three analytical continuation methods as introduced above. It includes four different analytical continuation solvers, namely \texttt{MaxEnt}, \texttt{StochAC}, \texttt{StochSK}, and \texttt{StochOM}. Just as their names suggested, the \texttt{MaxEnt} solver implements the maximum entropy method~\cite{JARRELL1996133}. The \texttt{StochAC} and \texttt{StochSK} solvers implement the K. S. D. Beach's algorithm~\cite{beach} and A. W. Sandvik's algorithm~\cite{PhysRevB.57.10287} of the stochastic analytical continuation, respectively. The \texttt{StochOM} solver implements the stochastic optimization method~\cite{PhysRevB.62.6317}. The ACFlow toolkit also provides a convenient library, which can be used to prepare and carry out analytical continuation calculations flexibly. The major features of the ACFlow toolkit are summarized in Table~\ref{tab:feature}.

\begin{table}[ht]
\centering
\begin{tabular}{l|l|l|l|l}
\hline
Features & \texttt{MaxEnt} & \texttt{StochAC} & \texttt{StochSK} & \texttt{StochOM} \\
\hline
Matrix-valued Green's function & Y & N & N & N \\
Imaginary time grid            & Y & Y & Y & Y \\
Matsubara frequency grid       & Y & Y & Y & Y \\
Linear mesh                    & Y & Y & Y & Y \\
Nonlinear mesh                 & Y & Y & Y & Y \\
Fermionic kernel               & Y & Y & Y & Y \\
Bosonic kernel                 & Y & Y & Y & Y \\
Self-defined model function    & Y & N & N & N \\
Constrained analytical continuation & N & Y & Y & Y \\
Regeneration of input data     & Y & Y & Y & Y \\
Kramers-Kronig transformation  & Y & Y & Y & Y \\
Parallel computing             & N & Y & Y & Y \\
Parallel tempering             & N & Y & N & N \\
Interactive mode               & Y & Y & Y & Y \\
Script mode                    & Y & Y & Y & Y \\
Standard mode                  & Y & Y & Y & Y \\
\hline
\end{tabular}
\caption{Major features of the ACFlow toolkit. \texttt{MaxEnt}, \texttt{StochAC}, \texttt{StochSK}, and \texttt{StochOM} are the four analytical continuation solvers as implemented in this toolkit. \label{tab:feature}}
\end{table}

In Table~\ref{tab:feature}, ``Y'' means yes while ``N'' means no. ``Interactive mode'', ``Script mode'', and ``Standard model'' are the three running modes supported by the ACFlow toolkit. We will introduce them in next section. The \texttt{MaxEnt} solver supports the ``historic'', ``classic'', ``bryan'', and ``chi2kink'' algorithms to determine the $\alpha$ parameter. The \texttt{StochAC} solver is only compatible with a flat model function, while the \texttt{StochSK} and \texttt{StochOM} solvers don't rely on any default model functions. The \texttt{StochOM} solver does not support analytical continuation of fermionic imaginary time Green's function for the moment. 

\subsection{Implementations}

The ACFlow toolkit is developed with pure Julia language. Thanks to powerful type system and multiple dispatch paradigm of the Julia language, the four different analytical continuation solvers are integrated into an united software architecture. Redundant codes are greatly reduced. It is quite easy to implement new analytical continuation solver or add new features to the existing solvers in the future. Distributed computing is a built-in feature of Julia. So, it is straightforward to realize parallel calculations in the ACFlow toolkit. Now except for the \texttt{MaxEnt} solver, all the other solvers are parallelized.

\begin{table}[ht]
\centering
\begin{tabular}{l|l}
\hline
Filename & Description \\
\hline
\texttt{ACFlow.jl} & Entry of the ACFlow module. \\
\texttt{maxent.jl} & Maximum entropy method. \\
\texttt{sac.jl}    & Stochastic analytical continuation (K. S. D. Beach's algorithm). \\
\texttt{san.jl}    & Stochastic analytical continuation (A. W. Sandvik's algorithm). \\
\texttt{som.jl}    & Stochastic optimization method. \\
\texttt{global.jl} & Numerical and physical constants. \\
\texttt{types.jl}  & Basic data structures and computational parameters. \\
\texttt{base.jl}   & Driver for analytical continuation simulation. \\
\texttt{inout.jl}  & Read input data and write calculated results. \\
\texttt{config.jl} & Parse configuration file and extract computational parameters. \\
\texttt{math.jl}   & Root finding, numerical integration, interpolation, Einstein summation, and curve fitting. \\
\texttt{util.jl}   & Some utility functions. \\
\texttt{mesh.jl}   & Meshes for spectral density. \\
\texttt{grid.jl}   & Grids for input data. \\
\texttt{model.jl}  & Default model functions. \\
\texttt{kernel.jl} & Kernel functions. \\
\hline
\end{tabular}
\caption{List of source codes of the ACFlow toolkit. \label{tab:source}}
\end{table}

The source codes of the ACFlow toolkit are placed in the \texttt{acflow/src} folder. Their functions are summarized in Table~\ref{tab:source}. The documentation of the ACFlow toolkit is written by using the Markdown language and \texttt{Documenter.jl} package. The source codes are placed in the \texttt{acflow/docs} folder. The users can build documentation by themselves. Please see section~\ref{sec:usage} for how to do that. Or they can read the latest documentation in the following website:
\begin{verbatim}
    https://huangli712.github.io/projects/acflow/index.html
\end{verbatim}    
Ten tests and four tutorials are also shipped with the ACFlow toolkit. Their source codes are placed in the \texttt{acflow/test} folder. See \texttt{acflow/test/test.md} and \texttt{acflow/test/tutor.md} for more details. The code repository of the ACFlow toolkit is:
\begin{verbatim}
    https://github.com/huangli712/ACFlow
\end{verbatim} 

\section{Getting started\label{sec:usage}}

In this section, we will discuss how to install and use the ACFlow toolkit.

\subsection{Installation}

It is an easy task to install the ACFlow toolkit. First, since it is written in pure Julia language, it is necessary to install the Julia runtime environment at first. The newest version of Julia is always preferred (version $>$ 1.60). Since the core codes only rely on Julia's built-in standard library, no the third-party packages are needed. Second, just download source codes of the ACFlow toolkit from its github repository. It should be a compressed file, such as \texttt{acflow.zip} or \texttt{acflow.tar.gz}. Please uncompress it in your favorite directory by using the following commands:
\begin{verbatim}
    $ unzip acflow.zip
\end{verbatim}
or
\begin{verbatim}
    $ tar xvfz acflow.tar.gz
\end{verbatim}
Third, the users have to declare a new environment variable \texttt{ACFLOW\_HOME}. Supposed that the root directory of the ACFLow toolkit is \texttt{/home/your\_home/acflow}, then \texttt{ACFLOW\_HOME} should be setup as follows:
\begin{verbatim}
    $ export ACFLOW_HOME=/home/your_home/acflow/src
\end{verbatim}
Finally, in order to generate the documentation, the users should type the following commands in the terminal:
\begin{verbatim}
    $ pwd
    /home/your_home/acflow
    $ cd docs
    $ julia make.jl
\end{verbatim}
After a few seconds, the documentation is built and saved in the \texttt{acflow/docs/build} directory if everything is OK. The home page of the documentation is \texttt{acflow/docs/build/index.html}. We can open it with any web browsers.   

\subsection{Run}

The ACFlow toolkit is designed to be flexible and easy-to-use. It provides three running modes to facilitate analytical continuation calculations, namely the interactive, script, and standard modes.  

\emph{Interactive mode}. With the ACFlow toolkit, the users can setup and carry out analytical continuation simulations interactively in Julia's REPL (Read-Eval-Print Loop) environment. For example,
\begin{verbatim}
    julia> push!(LOAD_PATH, ENV["ACFLOW_HOME"])
    julia> using ACFlow
    julia> setup_args("ac.toml")
    julia> read_param()
    julia> mesh, Aout, Gout = solve(read_data())
\end{verbatim}
Here, \texttt{ac.toml} is a configuration file, which contains essential computational parameters. The return values of the \texttt{solve()} function (i.e., \texttt{mesh}, \texttt{Aout}, and \texttt{Gout}) are mesh at real axis $\omega$, spectral density $A(\omega)$, and reproduced Green's function $\tilde{G}$, respectively. They can be further analyzed or visualized by the users.  

\emph{Script mode}. The core functionalities of the ACFlow toolkit are exposed to the users via a simple application programming interface. So, the users can write Julia scripts easily by themselves to perform analytical continuation simulations. A minimal Julia script (\texttt{acrun.jl}) is listed as follows:
\begin{verbatim}
    #!/usr/bin/env julia
    push!(LOAD_PATH, ENV["ACFLOW_HOME"])
    using ACFlow
    setup_args("ac.toml")
    read_param()
    mesh, Aout, Gout = solve(read_data())
\end{verbatim}
Of course, this script can be extended to finish complex tasks. In section~\ref{subsec:sigma}, a realistic example is provided to show how to complete an analytical continuation of Matsubara self-energy function via the script mode.              

\emph{Standard mode}. In the standard mode, the users have to prepare the input data manually. In addition, a configuration file must be provided. Supposed that the configuration file is \texttt{ac.toml}, then the analytical continuation calculation is launched as follows:
\begin{verbatim}
    $ /home/your_home/acflow/util/acrun.jl ac.toml
\end{verbatim}
or
\begin{verbatim}
    $ /home/your_home/acflow/util/Pacrun.jl ac.toml
\end{verbatim}
Noted that the \texttt{acrun.jl} script runs sequentially, while the \texttt{Pacrun.jl} script supports parallel and distributed computing. As we can conclude from the filename extension of configuration file (\texttt{ac.toml}), it adopts the TOML specification. The users may edit it with any text-based editors. Next we will introduce syntax and format of the input data files and configuration files. 

\subsection{Input files\label{subsec:input}}

The input files for the ACFlow toolkit can be divided into two groups: data files and configuration files. 

\emph{Data files}. The input data should be store in CSV-like text files. For imaginary time Green's function, the data file should contain three columns. They represent $\tau$, $\bar{G}(\tau)$, and standard deviation of $\bar{G}(\tau)$. For fermionic Matsubara Green's function, the data file should contain five columns. They represent $\omega_n$, Re$G(i\omega_n)$, Im$G(i\omega_n)$, standard deviation of Re$G(i\omega_n)$, and standard deviation of Im$G(i\omega_n)$. For bosonic correlation function $\chi(i\omega_n)$, the data file should contain four columns. They represent $\omega_n$, Re$\chi(i\omega_n)$, and standard deviation of Re$\chi(i\omega_n)$.

\emph{Configuration files}. The configuration file adopts the TOML format. It is used to setup the computational parameters. It consists of one or more blocks. Possible blocks (or sections) of the configuration file include \texttt{[BASE]}, \texttt{[MaxEnt]}, \texttt{[StochAC]}, \texttt{[StochSK]}, and \texttt{[StochOM]}. The \texttt{[BASE]} block is mandatory, while the other blocks are optional. A schematic configuration file (\texttt{ac.toml}) is listed as follows:
  
\begin{lstlisting}[language=TOML,
basicstyle=\ttfamily\small,
backgroundcolor=\color{yellow!10},
commentstyle=\color{olive!10!green},
keywordstyle=\color{purple}]
[BASE]
finput = "giw.data"
solver = "StochOM"
...

[MaxEnt]
method = "chi2kink"
...

[StochAC]
nfine  = 10000
...

[StochSK]
method = "chi2min"
...

[StochOM]
ntry   = 100000
...
\end{lstlisting}
In the \texttt{[BASE]} block, the analytical continuation problem is defined. The solver used to solve the problem must be assigned. The types of mesh, grid, default model function, and kernel function are also determined. The \texttt{[MaxEnt]}, \texttt{[StochAC]}, \texttt{[StochSK]}, and \texttt{[StochOM]} blocks are used to customize the corresponding analytical continuation solvers further. In Table~\ref{tab:base}-Table~\ref{tab:som}, all the possible input parameters for these blocks are collected and summarized. As for detailed explanations of these parameters, please refer to the user guide of the ACFlow toolkit. The uses can find it in the \texttt{acflow/docs} directory.   
  
\begin{table}[h]
\centering
\begin{tabular}{l|l|l|l}
\hline
\multicolumn{4}{c}{\texttt{[BASE]} block} \\
\hline
Parameter & Type & Default & Description \\
\hline
\texttt{finput}  & string  & ``green.data'' & Filename for input data. \\
\texttt{solver}  & string  & ``MaxEnt''     & Solver for the analytical continuation problem. \\
\texttt{ktype}   & string  & ``fermi''      & Type of kernel function. \\
\texttt{mtype}   & string  & ``flat''       & Type of default model function. \\
\texttt{grid}    & string  & ``ffreq''     & Grid for input data (imaginary axis). \\
\texttt{mesh}    & string  & ``linear''     & Mesh for output data (real axis). \\
\texttt{ngrid}   & integer & 10             & Number of grid points. \\
\texttt{nmesh}   & integer & 501            & Number of mesh points. \\
\texttt{wmax}    & float   & 5.0            & Right boundary (maximum value) of mesh. \\
\texttt{wmin}    & float   & -5.0           & Left boundary (minimum value) of mesh. \\
\texttt{beta}    & float   & 10.0           & Inverse temperature. \\
\texttt{offdiag} & bool    & false          & Treat the off-diagonal part of matrix-valued function? \\
\texttt{pmodel}  & array   & N/A            & Additional parameters for customizing the default model. \\
\texttt{pmesh}   & array   & N/A            & Additional parameters for customizing the mesh. \\
\texttt{exclude} & array   & N/A            & Restriction of energy range of the spectrum. \\
\hline
\end{tabular}
\caption{Possible parameters for the \texttt{[BASE]} block.\label{tab:base}}
\end{table}

\begin{table}[h]
\centering
\begin{tabular}{l|l|l|l}
\hline
\multicolumn{4}{c}{\texttt{[MaxEnt]} block} \\
\hline
Parameter & Type & Default & Description \\
\hline
\texttt{method} & string  & ``chi2kink''& How to determine the optimized $\alpha$ parameter? \\
\texttt{nalph}  & integer & 12          & Total number of the chosen $\alpha$ parameters. \\
\texttt{alpha}  & float   & 1e9         & Starting value for the $\alpha$ parameter. \\
\texttt{ratio}  & float   & 10.0        & Scaling factor for the $\alpha$ parameter. \\
\texttt{blur}   & float   & -1.0        & Shall we preblur the kernel and spectrum?\\
\hline
\end{tabular}
\caption{Possible input parameters for the \texttt{[MaxEnt]} block, which are used to setup the solver based on the maximum entropy method~\cite{JARRELL1996133,PhysRevB.44.6011}. \label{tab:maxent}}
\end{table}

\begin{table}[h]
\centering
\begin{tabular}{l|l|l|l}
\hline
\multicolumn{4}{c}{\texttt{[StochAC]} block} \\
\hline
Parameter & Type & Default & Description \\
\hline
\texttt{nfine}  & integer & 10000       & Number of points of a very fine linear mesh. \\
\texttt{ngamm}  & integer & 512         & Number of $\delta$ functions. \\
\texttt{nwarm}  & integer & 4000        & Number of Monte Carlo thermalization steps. \\
\texttt{nstep}  & integer & 4000000     & Number of Monte Carlo sweeping steps. \\
\texttt{ndump}  & integer & 40000       & Intervals for monitoring Monte Carlo sweeps. \\
\texttt{nalph}  & integer & 20          & Total number of the chosen $\alpha$ parameters. \\
\texttt{alpha}  & float   & 1.0         & Starting value for the $\alpha$ parameter. \\
\texttt{ratio}  & float   & 1.2         & Scaling factor for the $\alpha$ parameter. \\
\hline
\multicolumn{4}{c}{\texttt{[StochSK]} block} \\
\hline
Parameter & Type & Default & Description \\
\hline
\texttt{method} & string  & ``chi2min'' & How to determine the optimized $\Theta$ parameter? \\
\texttt{nfine}  & integer & 100000      & Number of points of a very fine linear mesh. \\
\texttt{ngamm}  & integer & 1000        & Number of $\delta$ functions. \\
\texttt{nwarm}  & integer & 1000        & Number of Monte Carlo thermalization steps. \\
\texttt{nstep}  & integer & 20000       & Number of Monte Carlo sweeping steps. \\
\texttt{ndump}  & integer & 200         & Intervals for monitoring Monte Carlo sweeps. \\
\texttt{retry}  & integer & 10          & How often to recalculate the goodness-of-fit function. \\
\texttt{theta}  & float   & 1e6         & Starting value for the $\Theta$ parameter. \\
\texttt{ratio}  & float   & 0.9         & Scaling factor for the $\Theta$ parameter. \\
\hline
\end{tabular}
\caption{Possible input parameters for the \texttt{[StochAC]} and \texttt{[StochSK]} blocks, which are used to setup the two solvers based on the stochastic analytical continuation (Beach's and Sandvik's algorithms)~\cite{PhysRevB.57.10287,beach}. \label{tab:sac}}
\end{table}

\begin{table}[h]
\centering
\begin{tabular}{l|l|l|l}
\hline
\multicolumn{4}{c}{\texttt{[StochOM]} block} \\
\hline
Parameter & Type & Default & Description \\
\hline
\texttt{ntry}   & integer & 2000        & Number of attempts to figure out the solution. \\
\texttt{nstep}  & integer & 1000        & Number of Monte Carlo steps per try. \\
\texttt{nbox}   & integer & 100         & Number of boxes to construct the spectrum. \\
\texttt{sbox}   & float   & 0.005       & Minimum area of the randomly generated rectangles. \\
\texttt{wbox}   & float   & 0.02        & Minimum width of the randomly generated rectangles. \\
\texttt{norm}   & float   & -1.0        & Is the norm calculated? \\
\hline
\end{tabular}
\caption{Possible input parameters for the \texttt{[StochOM]} block, which are used to setup the solver based on the stochastic optimization method~\cite{PhysRevB.62.6317}. \label{tab:som}}
\end{table}

\subsection{Output files\label{subsec:output}}

Once the analytical continuation simulation is finished, the final spectral function $A(\omega)$ is outputted to \texttt{Aout.data}. As is shown in Eq.~(\ref{eq:ImG}), $A(\omega)$ is equivalent to the imaginary part of real frequency Green's function Im$G(\omega)$. Then the ACFlow toolkit will automatically calculate the corresponding real part Re$G(\omega)$ via the Kramers-Kronig transformation [see Eq.~(\ref{eq:kk})]. The full Green's function at real axis $G(\omega)$ is stored in \texttt{Gout.data}. The spectral function is also used to reconstruct the imaginary time or Matsubara Green's functions [$\tilde{G}(\tau)$ or $\tilde{G}(i\omega_n)$], which is stored in \texttt{repr.data}. Besides the three output files, the ACFlow toolkit will generate quite a few output files, which can be used to analyze and diagnose the calculated results. All of the possible output files of the ACFlow toolkit are collected and explained in Table~\ref{tab:output}. 
 
\begin{table}[ht]
\centering
\begin{tabular}{l|l}
\hline
Filename & Description \\
\hline
\texttt{Aout.data} & Final spectral function $A(\omega)$. \\
\texttt{Gout.data} & Full Green's function at real axis $G(\omega)$. \\
\texttt{repr.data} & Reproduced Green's function $\tilde{G}$ at imaginary time or frequency axis. \\
\texttt{model.data} & Default model function $m(\omega)$. \\
\texttt{chi2.data} & $\log_{10}(\chi^2)$ vs $\log_{10}(\alpha)$. \\
\texttt{prob.data} & $P[\alpha | \bar{G}]$ vs $\alpha$ for the \texttt{MaxEnt} solver (bryan algorithm). \\
\texttt{Aout.data.alpha}\_$i$ & $\alpha$-resolved spectral function $A_{\alpha}(\omega)$ for the \texttt{StochAC} solver. \\
\texttt{hamil.data} & $U(\alpha)$ vs $\alpha$ for the \texttt{StochAC} solver. \\
\texttt{goodness.dat} & $\log_{10}(\chi^2)$ vs $\log_{10}(\Theta)$ for the \texttt{StochSK} solver. \\
\texttt{stat.data} & Monte Carlo statistical information for stochastic sampling method. \\
\hline
\end{tabular}
\caption{Possible output files of the ACFlow toolkit. \label{tab:output}}
\end{table}

\section{Examples\label{sec:examples}}

In order to demonstrate usefulness of the ACFlow toolkit, four examples are illustrated in this section. These examples cover typical application scenarios of the ACFlow toolkit, including analytical continuations of Matsubara self-energy function, Matsubara Green's function, imaginary time Green's function, and current-current correlation function within the script mode or standard mode. All of the necessary source codes and data files, which can be used to reproduce the results as shown in this section, are placed in the \texttt{/home/your\_home/acflow/test/T*} folders.  

\subsection{Matsubara self-energy function\label{subsec:sigma}}

Now let us consider the following single-band Hubbard model on a Bethe lattice at first:
\begin{equation}
H = -t \sum_{\langle ij \rangle \sigma} c^{\dagger}_{i\sigma}c_{j\sigma}
 - \mu \sum_i n_i + U \sum_i n_{i\uparrow} n_{i\downarrow},
\end{equation}
where $t$ is the hopping parameter, $\mu$ is the chemical potential, $U$ is the Coulomb interaction, $n$ is the occupation number, $\sigma$ denotes the spin, $i$ and $j$ are site indices. This model is solved by using the dynamical mean-field theory (dubbed DMFT)~\cite{RevModPhys.68.13} with the hybridization expansion continuous-time quantum Monte Carlo solver (dubbed CT-HYB)~\cite{RevModPhys.83.349} as implemented in the $i$QIST package~\cite{HUANG2015140,HUANG2017423}. The parameters used in the DMFT + CT-HYB calculation are $t = 0.5$, $U = 2.0$, $\mu = 1.0$, and $\beta = 10.0$. Once the DMFT self-consistent calculation is finished, the Matsubara self-energy function $\Sigma(i\omega_n)$ is obtained. We are going to convert it to real frequency self-energy function $\Sigma(\omega)$. The data of Matsubara self-energy function $\Sigma(i\omega_n)$ have been preprocessed and stored in \texttt{siw.data}. This file contains five columns, which are used to record the Matsubara frequency $\omega_n$, Re$\Sigma(i\omega_n)$, Im$\Sigma(i\omega_n)$, error bar of Re$\Sigma(i\omega_n)$, error bar of Im$\Sigma(i\omega_n)$, respectively. Only the first twenty Matsubara frequency points are kept, because the high-frequency data are somewhat noisy.

The purpose of this example is to demonstrate usage of the \texttt{MaxEnt} solver and the script mode of the ACFlow toolkit. Next we will explain the key steps in detail. As for the complete Julia script, please refer to \texttt{sigma.jl} and \texttt{gendata.jl} in the \texttt{/home/your\_home/acflow/test/T01/} folder.   

First, we have to load the essential Julia packages. Both the \texttt{DelimitedFiles} and \texttt{Printf} packages belong to Julia's standard library. They are used to read input data and write calculated results, respectively.  

\begin{lstlisting}[language=Julia,
basicstyle=\ttfamily\small,
backgroundcolor=\color{yellow!10},
commentstyle=\color{olive!10!green},
keywordstyle=\color{purple}]
#!/usr/bin/env julia

push!(LOAD_PATH, ENV["ACFLOW_HOME"])

using DelimitedFiles
using Printf
using ACFlow

welcome() # Print welcome message only
\end{lstlisting}

Next, the data of Matsubara self-energy function are read from \texttt{siw.data}. The Hartree term $\Sigma_{H}$ should be subtracted from its real part:
\begin{equation}
\Sigma(i\omega_n) \to \Sigma(i\omega_n) - \Sigma_{H}.
\end{equation}
Note that $\Sigma_{H}$ is approximately equal to the asymptotic value of real part of $\Sigma(i\omega_n)$ when $n$ goes to infinite.   
 
\begin{lstlisting}[language=Julia,
basicstyle=\ttfamily\small,
backgroundcolor=\color{yellow!10},
commentstyle=\color{olive!10!green},
keywordstyle=\color{purple}]
# Deal with self-energy function
#
# Read self-energy function
dlm = readdlm("siw.data")
#
# Get grid
grid = dlm[:,1]
#
# Get self-energy function
Sinp = dlm[:,2] + im * dlm[:,3] # Value
Serr = dlm[:,4] + im * dlm[:,5] # Error bar
#
# Subtract hartree term
Sh = 1.0
@. Sinp = Sinp - Sh
\end{lstlisting}

Next, the computational parameters are encapsulated into two dictionaries. The dictionary \texttt{B} is for the \texttt{[BASE]} block, while the dictionary \texttt{S} is for the \texttt{MaxEnt} solver. Then the \texttt{setup\_param()} function is called, so that these parameters take effect. Here, the \texttt{MatEnt} solver~\cite{JARRELL1996133,PhysRevB.44.6011} is employed to tackle the analytical continuation problem. But the other stochastic sampling solvers are also applicable. The default model function is gaussian. The mesh for spectral density is non-uniform (A tangent mesh). The number of used $\alpha$ parameters is 15, and the optimal $\alpha$ parameter is determined by the $\chi^2$kink algorithm~\cite{PhysRevE.94.023303}. 

\begin{lstlisting}[language=Julia,
basicstyle=\ttfamily\small,
backgroundcolor=\color{yellow!10},
commentstyle=\color{olive!10!green},
keywordstyle=\color{purple}]
# Setup parameters
#
# For [BASE] block
# See types.jl/_PBASE for default setup
B = Dict{String,Any}(
    "solver" => "MaxEnt",  # Choose MaxEnt solver
    "mtype"  => "gauss",   # Default model function
    "mesh"   => "tangent", # Mesh for spectral density
    "ngrid"  => 20,        # Number of input points
    "nmesh"  => 801,       # Number of output points
    "wmax"   => 8.0,       # Right boundary of mesh
    "wmin"   => -8.0,      # Left boundary of mesh
    "beta"   => 10.0,      # Inverse temperature
)
#
# For [MaxEnt] block
# See types.jl/_PMaxEnt for default setup
S = Dict{String,Any}(
    "nalph"  => 15,        # Number of alpha
    "alpha"  => 1e12,      # Starting value of alpha
    "blur"   => -1.0,      # Enable preblur or not
)
#
# Let the parameters take effect
setup_param(B, S)
\end{lstlisting}

It is quite easy to start the analytical continuation calculation. Just call the \texttt{solve()} function and pass the grid, input data, and error bar data to it. The return values of this function call are real frequency mesh, spectral density, and reconstructed Matsubara self-energy function. 

\begin{lstlisting}[language=Julia,
basicstyle=\ttfamily\small,
backgroundcolor=\color{yellow!10},
commentstyle=\color{olive!10!green},
keywordstyle=\color{purple}]
# Call the solver
mesh, Aout, Sout = solve(grid, Sinp, Serr)
\end{lstlisting}

Finally, the real frequency self-energy function must be supplemented with the Hartree term, and then the final results are written into \texttt{sigma.data}.   
   
\begin{lstlisting}[language=Julia,
basicstyle=\ttfamily\small,
backgroundcolor=\color{yellow!10},
commentstyle=\color{olive!10!green},
keywordstyle=\color{purple}]
# Calculate final self-energy function on real axis
#
# Add hartree term
@. Sout = Sout + Sh
#
# Write self-energy function to sigma.data
open("sigma.data", "w") do fout
    for i in eachindex(mesh)
        z = Sout[i]
        @printf(fout, "%20.16f %20.16f %20.16f\n",
                mesh[i], real(z), imag(z))
    end
end
\end{lstlisting}

\begin{figure}[ht]
\centering
\includegraphics[width=\textwidth]{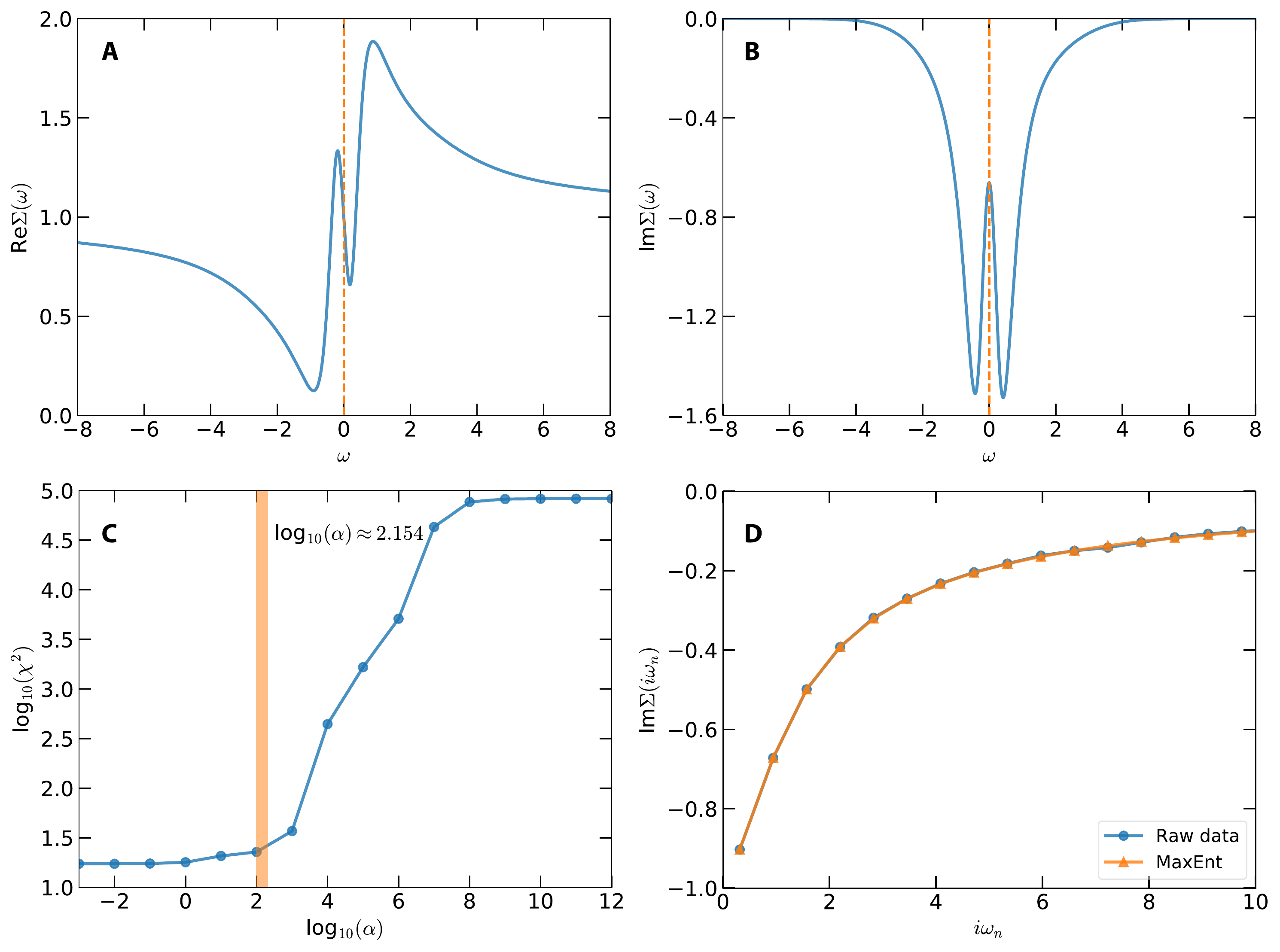}
\caption{Analytical continuation of Matsubara self-energy function by using the maximum entropy method. (a) Real part of real frequency self-energy function. (b) Imaginary part of real frequency self-energy function. (c) $\chi^{2}$ as a function of $\alpha$. The vertical bar indicates the optimal $\alpha$ parameter chosen by the $\chi^2$kink algorithm. (d) Reproduced and original data for imaginary part of the Matsubara self-energy functions. \label{fig:sig}}
\end{figure}

The calculated results are displayed in Fig.~\ref{fig:sig}. Fig.~{\ref{fig:sig}}(a) and (b) show the real and imaginary parts of the real frequency self-energy function, respectively. Near the Fermi level, Re$\Sigma(\omega)$ exhibits quasi-linear behavior, with which the quasiparticle weight $Z$ and effective mass of electron $m^*$ can be easily evaluated~\cite{RevModPhys.68.13}. As for the imaginary part, Im$\Sigma(0)$ is finite, which indicates that the electron-electron scattering is not trivial. Fig.~\ref{fig:sig}(c) shows the $\alpha$-dependent $\chi^{2}$. The vertical bar in this figure indicates the optimal $\alpha$ is around 10$^{2.154}$. In Fig.~\ref{fig:sig}(d), the reproduced and raw Matsubara self-energy functions are compared. It is apparent that they are consistent with each other.

\subsection{Matsubara Green's function\label{subsec:green}}

The purpose of the second example is to treat the Matsubara Green's function by using the \texttt{StochOM} solver.     

At first, please consider the following spectral density with two gaussian peaks:
\begin{equation}
A(\omega) = 
A_1 \exp\left[\frac{-(\omega - \epsilon_1)^2}{2 \Gamma^2_1}\right] +
A_2 \exp\left[\frac{-(\omega - \epsilon_2)^2}{2 \Gamma^2_2}\right],
\end{equation}
with $A_1 = 1.0$, $A_2 = 0.3$, $\epsilon_1 = 0.5$, $\epsilon_2 = -2.5$, $\Gamma_1 = 0.2$, and $\Gamma_2 = 0.8$. Then the Matsubara Green's function $G(i\omega_n)$ is evaluated by using Eq.~(\ref{eq:kernel_w}) with $\beta = 10.0$. Random noises, generated by the formula $0.0001 r_1 \exp(i 2\pi r_2 )$ where $r_1$ and $r_2$ are pseudo random numbers in ($0.0 < r_1$, $r_2 < 1.0$), are added to $G(i\omega_n)$. The error bar of $G(i\omega_n)$ is fixed to 1e-4. The generated data for $G(i\omega_n)$ are written in \texttt{giw.data}.  

Next, we are going to use the standard mode, such that a configure file (\texttt{ac.toml}) must be prepared. It is listed as follows. Since the \texttt{StochOM} solver is chosen, the \texttt{[BASE]} and \texttt{[StochOM]} blocks must be present. 

\begin{lstlisting}[language=TOML,
basicstyle=\ttfamily\small,
backgroundcolor=\color{yellow!10},
commentstyle=\color{olive!10!green},
keywordstyle=\color{purple}]
[BASE]
finput = "giw.data"
solver = "StochOM"
ktype  = "fermi"
mtype  = "flat"
grid   = "ffreq"
mesh   = "linear"
ngrid  = 10
nmesh  = 501
wmax   = 5.0
wmin   = -5.0
beta   = 10.0
offdiag = false

[StochOM]
ntry  = 100000
nstep = 1000
nbox  = 100
sbox  = 0.005
wbox  = 0.02
norm  = -1.0
\end{lstlisting}

\begin{figure}[ht]
\centering
\includegraphics[width=\textwidth]{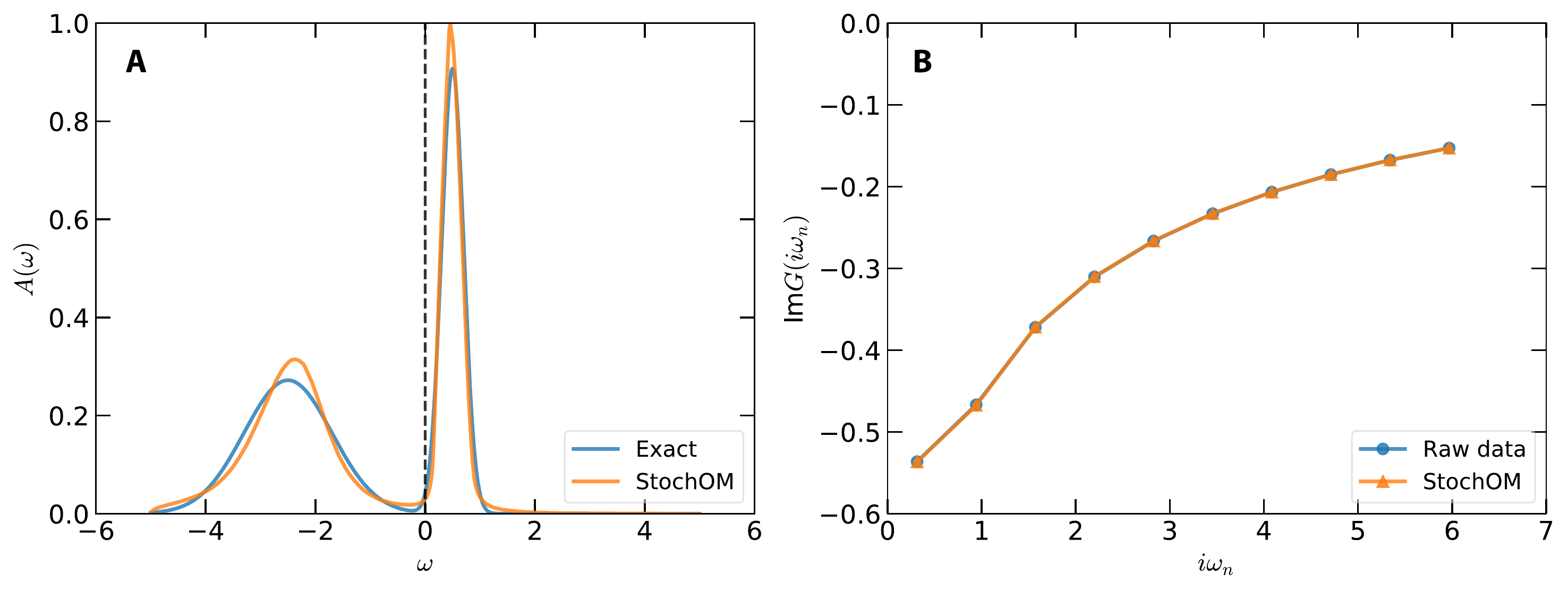}
\caption{Analytical continuation of Matsubara Green's function by using the stochastic optimization method. (a) Simulated and exact spectral functions. (b) Reconstructed and synthetic Matsubara Green's functions. Only the imaginary parts are presented in this figure. \label{fig:giw}}
\end{figure}

Then we use the \texttt{acrun.jl} or \texttt{Pacrun.jl} script to perform analytical continuation simulation. The calculated results are shown in Fig.~\ref{fig:giw}. As is seen in Fig.~\ref{fig:giw}(a), both the sharp peak around 0.5~eV and the broad peak around -2.5~eV are correctly reproduced by the \texttt{StochOM} solver. In Fig.~\ref{fig:giw}(b), the reconstructed Matsubara Green's function agrees quite well with the raw input data.

\subsection{Imaginary time Green's function\label{subsec:gtime}}

In this example, analytical continuation of imaginary time Green's function will be tested. Note that this example is borrowed from Reference~\cite{beach} directly.

The exact spectral function reads:
\begin{equation}
A(\omega) =
\begin{cases}
\frac{1}{W} \frac{|\omega|}{\sqrt{\omega^2 - \Delta^2}},~\quad & \text{if}~\Delta < |\omega| < W/2. \\
0, & \text{otherwise}.
\end{cases} 
\end{equation}
Here, $W$ denotes bandwidth, and $\Delta$ is used to control size of the energy gap. Let $W = 6$ and $2\Delta = 1$. This spectrum should exhibit flat shoulders, steep peaks, and sharp gap edges at the same time. Actually, it is a typical spectrum of a BCS superconductor. 

First, the imaginary time Green's function $G(\tau)$ is generated using Eq.~(\ref{eq:kernel_t}). Then a normally-distributed random noise is add to $G(\tau)$. Maximum amplitude of the noise is 1e-4. The error bar of $G(\tau)$ is fixed to 1e-3. The data are written in \texttt{gtau.data}.

Next, we try to prepare the configure file (\texttt{ac.toml}). In this case, we would like to benchmark the \texttt{StochAC} solver, so the \texttt{solver} parameter is set to ``StochAC'' and the \texttt{grid} parameter is set to ``ftime''. Furthermore, the \texttt{exclude} parameter is enabled to impose some \emph{a priori} constraints to the spectrum. The full \texttt{ac.toml} is listed as follows:

\begin{lstlisting}[language=TOML,
basicstyle=\ttfamily\small,
backgroundcolor=\color{yellow!10},
commentstyle=\color{olive!10!green},
keywordstyle=\color{purple}]
[BASE]
finput = "giw.data"
solver = "MaxEnt"
ktype  = "fermi"
mtype  = "flat"
grid   = "ffreq"
mesh   = "linear"
ngrid  = 10
nmesh  = 501
wmax   = 5.0
wmin   = -5.0
beta   = 10.0
offdiag = false
exclude = [[-5.0,-3.0], [-0.5,0.5], [3.0,5.0]]

[StochAC]
nfine = 10000
ngamm = 512
nwarm = 4000
nstep = 10000000
ndump = 40000
nalph = 40
alpha = 1.00
ratio = 1.20
\end{lstlisting}

We perform analytical continuation simulation by using the \texttt{acrun.jl} or \texttt{Pacrun.jl} script. In order to obtain smooth spectral density, it is useful to increase number of $\delta$ functions (See \texttt{ngamm} parameter) and number of Monte Carlo sampling steps (See \texttt{nstep} parameter).     

\begin{figure}[ht]
\centering
\includegraphics[width=\textwidth]{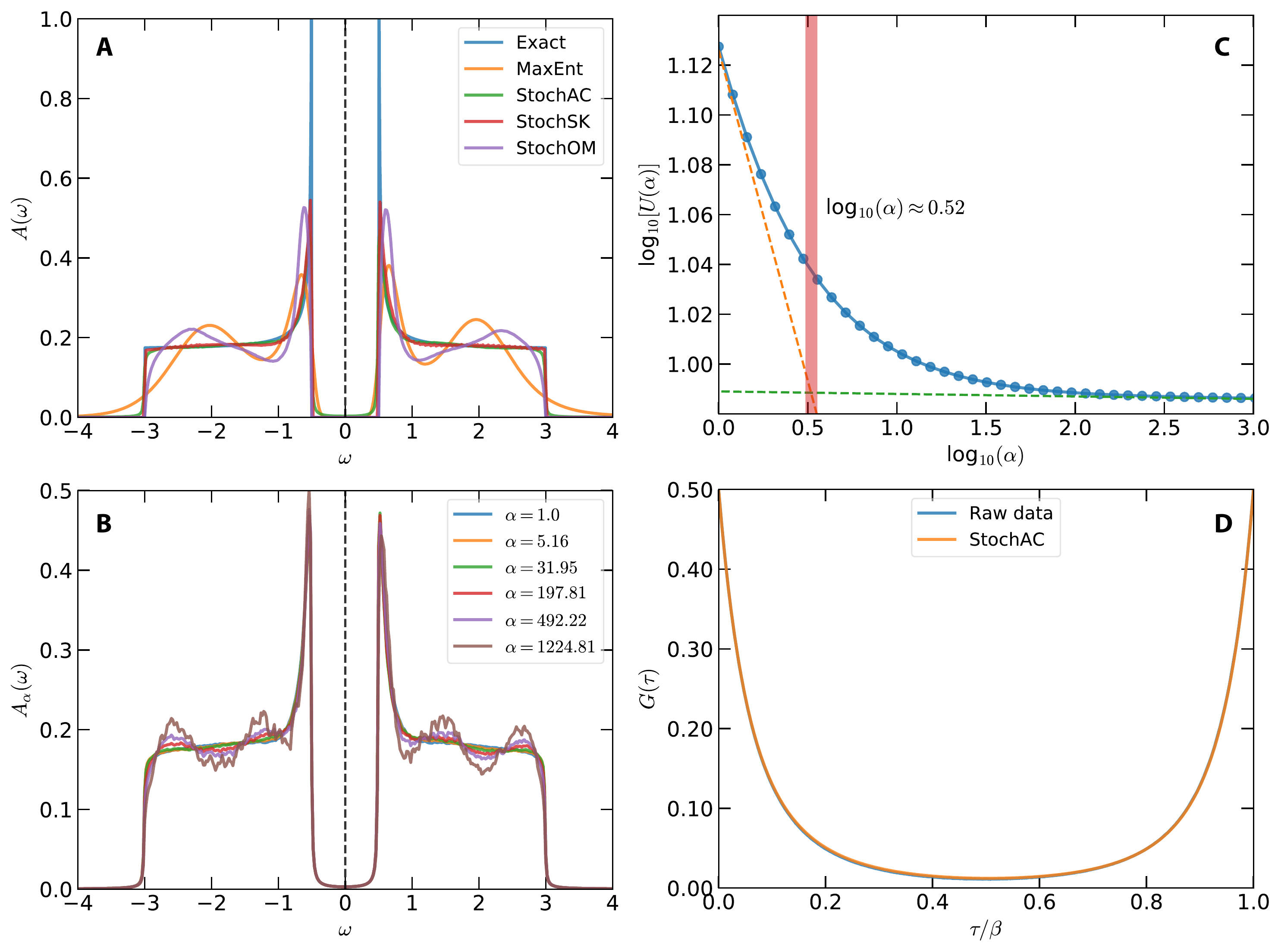}
\caption{Analytical continuation of imaginary time Green's function by using the stochastic analytical continuation (Beach's algorithm). (a) Simulated and exact spectral functions. (b) $\alpha$-dependent spectral functions. (c) Internal energy $U$ as a function of $\alpha$. The vertical bar indicates the optimal $\alpha$ parameter. (d) Simulated and exact imaginary time Green's functions. \label{fig:gtau}}
\end{figure}
 
Figure~\ref{fig:gtau} shows the calculated results. In Fig.~\ref{fig:gtau}(a), the exact spectral function is compared with the simulated spectrum. Note that besides the \texttt{StochAC} solver, the other three solvers are also tested. Their results are also plotted in this figure for a direct comparison. It is remarkable that the \texttt{StochAC} and \texttt{StochSK} solvers do a superior job of modelling the spectrum. The major characteristics of the spectrum, including flat regions, steep peaks, and sharp gap edges, are well captured by the two solvers. Especially, we have finished more tests without any constraints on the spectral density. The gap in the spectrum can be reproduced as well. On the other hand, the spectra obtained by the \texttt{MaxEnt} and \texttt{StochOM} solvers are much too smooth, and show extra shoulder peaks around $\pm$2.0. Figure~\ref{fig:gtau}(b) shows $\alpha$-resolved spectral functions $A_{\alpha}(\omega)$ for selected $\alpha$ parameters. Fluctuation in the flat regions of the calculated spectral density grows when $\alpha$ increases. Figure~\ref{fig:gtau}(c) shows internal energy $U$ as a function of $\alpha$. From this figure, the critical $\alpha$ is estimated, which is indicated by the vertical bar. Finally, the reproduced Green's function $\tilde{G}(\tau)$ agrees quite well with the raw input data, as is shown in Fig.~\ref{fig:gtau}(d).     

\subsection{Current-current correlation function\label{subsec:optic}}

The previous three examples only involve fermionic correlators. How about bosonic correlation functions? In this example, we will demonstrate how to perform analytical continuation simulation for a typical bosonic correlator, the current-current correlation function $\Pi(\tau)$, to obtain the optical conductivity $\sigma(\omega)$. Note that this example is taken from Reference~\cite{PhysRevB.82.165125} directly. 

The exact optical conductivity $\sigma(\omega)$ reads:
\begin{equation}
\label{eq:optic}
\sigma(\omega) = 
\left\{
\frac{W_1}{1 + (\omega/\Gamma_1)^2} + 
\frac{W_2}{1 + [(\omega - \epsilon)/\Gamma_2]^2} +
\frac{W_2}{1 + [(\omega + \epsilon)/\Gamma_2]^2}
\right\}
\frac{1}{1 + (\omega/\Gamma_3)^6},
\end{equation}
where $W_1 = 0.3$, $W_2 = 0.2$, $\Gamma_1 = 0.3$, $\Gamma_2 = 1.2$, $\Gamma_3 = 4.0$, and $\epsilon = 3.0$. The current-current correlation function $\Pi(\tau)$ can be evaluated from $\sigma(\omega)$ by using the following equation:
\begin{equation}
\label{eq:current}
\Pi(\tau) = \int^{\infty}_{-\infty} K(\tau,\omega) \sigma(\omega)~d\omega,
\end{equation}
where the kernel function $K(\tau,\omega)$ is different from Eq.~(\ref{eq:ktau}). It reads:
\begin{equation}
\label{eq:Koptic}
K(\tau,\omega) = \frac{1}{\pi} \frac{\omega e^{-\tau\omega}}{1- e^{-\beta\omega}}.
\end{equation}
In this case, $\beta$ is fixed to be 20.0. 

At first, we use Eq.~(\ref{eq:optic}) $\sim$ Eq.~(\ref{eq:Koptic}) to prepare $\Pi(\tau)$. A normally-distributed random noise is add to $\Pi(\tau)$. Maximum amplitude of the noise is 1e-4. The error bar of $\Pi(\tau)$ is fixed to 1e-4. The data of $\Pi(\tau)$ are written in \texttt{chit.data}. 

Next, we conduct analytical continuation simulation as usual. The used configuration file is attached as follows. Here, the \texttt{StochSK} solver is adopted, so the \texttt{solver} parameter is ``StochSK'' and the \texttt{grid} parameter is ``btime''. And the Shao-Sandvik algorithm~\cite{PhysRevX.7.041072} is applied to seek optimal $\Theta$, so the \texttt{method} parameter is ``chi2min''. The users can further increase the values of \texttt{nfine}, \texttt{ngamm}, and \texttt{nstep} parameters to improve computational accuracy. 
 
\begin{lstlisting}[language=TOML,
basicstyle=\ttfamily\small,
backgroundcolor=\color{yellow!10},
commentstyle=\color{olive!10!green},
keywordstyle=\color{purple}]
[BASE]
finput = "chit.data"
solver = "StochSK"
ktype  = "bsymm"
mtype  = "flat"
grid   = "btime"
mesh   = "linear"
ngrid  = 501
nmesh  = 801
wmax   = 8.0
wmin   = 0.0
beta   = 20.0
offdiag = false

[StochSK]
method = "chi2min"
nfine = 40000
ngamm = 1000
nwarm = 1000
nstep = 1000000
ndump = 200
retry = 10
theta = 1e+6
ratio = 0.90
\end{lstlisting}

\begin{figure}[ht]
\centering
\includegraphics[width=\textwidth]{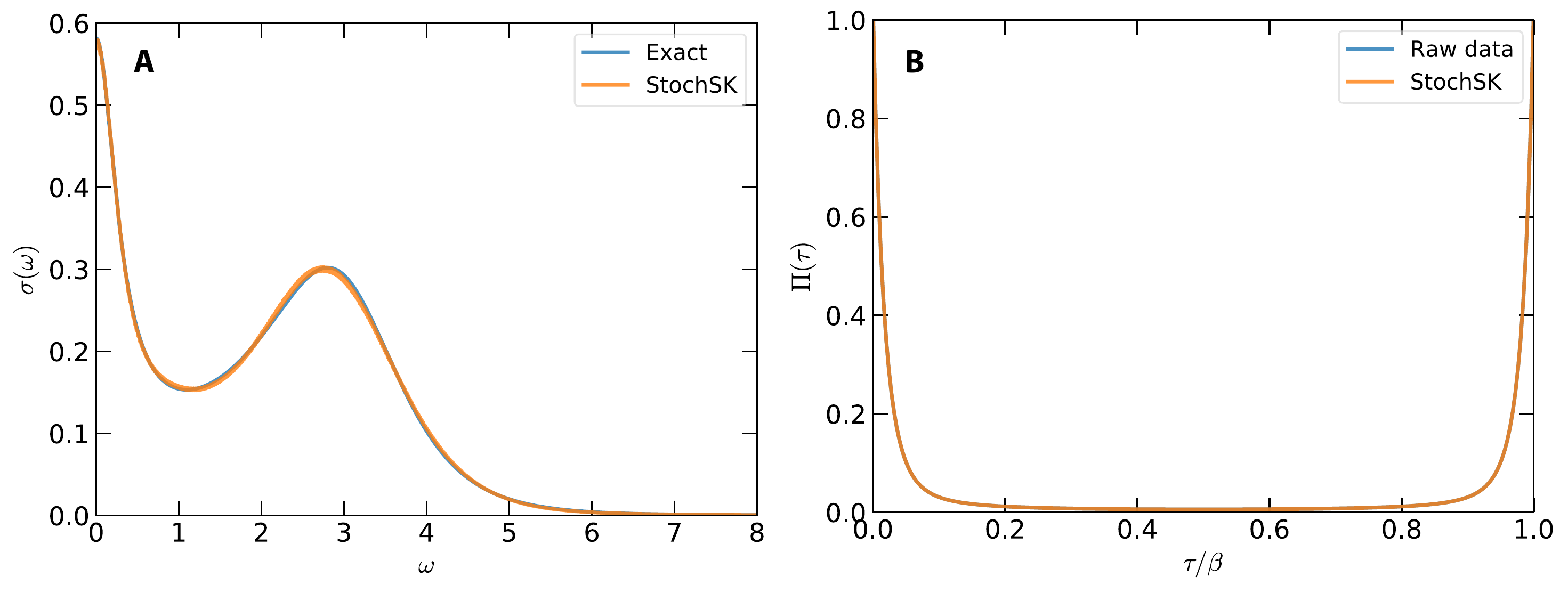}
\caption{Analytical continuation of current-current correlation function by using the stochastic analytical continuation (Sandvik's algorithm). (a) Simulated and exact optical conductivities $\sigma(\omega)$. (b) Simulated and exact current-current correlation functions $\Pi(\tau)$. \label{fig:optic}}
\end{figure}

The calculated results are illustrated in Fig.~\ref{fig:optic}. From Fig.~\ref{fig:optic}(a), it is clear that the main features of optical conductivity are successfully captured by the \texttt{StochSK} solver. Both the sharp Drude peak at $\omega = 0$ and a broad satellite peak around $\omega = 3.0$ are well reproduced. As is seen in Fig.~\ref{fig:optic}(b), the reconstructed $\tilde{\Pi}(\tau)$ coincides with the original $\Pi(\tau)$. 

\section{Concluding remarks\label{sec:outlook}}

In this paper, a full-fledged analytical continuation toolkit named ACFlow is presented. It implements several primary analytical continuation methods, including the maximum entropy method, stochastic analytical continuation, and stochastic optimization method. It provides quite a few validation and diagnostic tools. It can be used with great flexibility for the analytical continuations of arbitrary fermionic and bosonic correlation functions generated by finite-temperature quantum Monte Carlo simulations.

Note that analytical continuation problem is a hotspot in computational physics and many-body physics all the time. Many efforts have been devoted to solve it in recent years. Noticeable achievements include maximum quantum entropy method~\cite{PhysRevB.98.205102}, Nevanlinna analytical continuation~\cite{PhysRevLett.126.056402,PhysRevB.104.165111}, blocked-mode sampling and grid point sampling in stochastic analytical continuation~\cite{PhysRevB.101.085111,PhysRevB.102.035114}, constrained stochastic analytical continuation~\cite{PhysRevE.94.063308,Shao:2022yez}, machine learning assisted analytical continuation~\cite{PhysRevLett.124.056401,PhysRevB.98.245101}, and so on. We would like to incorporate these new progresses into the ACFlow toolkit in the near future.      

\vspace{10pt}

\noindent \textbf{Declaration of competing interest}

\vspace{5pt}
The authors declare that they have no known competing financial interests or personal relationships that could have appeared to influence the work reported in this paper.\\

\noindent \textbf{Data availability}

\vspace{5pt}
The data that support the findings of this study will be made available upon reasonable requests to the corresponding author.\\

\noindent \textbf{Acknowledgement}

\vspace{5pt}
This work is supported by the CAEP Foundation (under Grant No.~CX100000) and the National Natural Science Foundation of China (under Grants No.~11874329 and No.~11934020).\\
 
\bibliographystyle{elsarticle-num}
\bibliography{acflow}

\end{document}